\documentclass[aps,pra,twocolumn,nofootinbib,10pt]{revtex4-2}

\usepackage{cancel}
\usepackage{footnotebackref}
\usepackage{xcolor}
\usepackage{footmisc}
\usepackage{subcaption}
\usepackage{comment}

\usepackage{ragged2e}
\usepackage{amsmath, amssymb, booktabs}
\usepackage{graphicx}
\usepackage{dcolumn}
\usepackage{bm}
\usepackage{physics}
\usepackage{hyperref}

\begin{document}

\preprint{APS/123-QED}

\title{Magnetic noise in macroscopic quantum spatial superposition induced by inverted harmonic oscillator potential} 

\author{Sneha Narasimha Moorthy$^{1, \,2}$}
\author{Anupam Mazumdar$^{3}$}
\affiliation{ 
$^{1}$School of Physical Sciences, National Institute of Science Education and Research, Jatni 752050, India\\
$^{2}$Homi Bhabha National Institute, Training School Complex, Anushaktinagar, Mumbai 400094, India\\
$^{3}$ Van Swinderen Institute, University of Groningen, 9747 AG Groningen, The Netherlands\\}

\date{September 1, 2025}

\begin{abstract}
{We investigate a Stern-Gerlach type matter-wave interferometer where an inhomogeneous magnetic field couples to an embedded spin in a nanoparticle to create spatial superpositions. Employing a sequence of harmonic and inverted harmonic oscillator potentials created by external magnetic fields, we aim to enhance the one-dimensional superposition of a nanodiamond with mass $\sim 10^{-15}$ kg to $\sim 1 \mu$m. However, random fluctuations of the magnetic field stochastically perturbs the interferometer paths and induce dephasing. We quantitatively estimate the susceptibility of the interferometer to white noise arising from magnetic-field fluctuations. Constraining the dephasing rate \(\Gamma\) to be low enough that the final coherence \(e^{-\Gamma \tau}\leq 0.1\) (where \(\tau\) is the experimental time duration), we obtain the following bounds on the noise to signal ratios: $\delta \eta_\text{IHP}/\eta_\text{IHP}\lesssim 10^{-13}$, where $\eta_\text{IHP}$ is the magnetic field curvature that gives rise to the inverted harmonic potential, and $\delta \eta_\text{HP}/\eta_\text{HP}\lesssim 10^{-6}$, where $\eta_\text{HP}$ is the linear magnetic field gradient that gives rise to the harmonic potential. For such tiny fluctuations, we demonstrate that the Humpty-Dumpty problem arising from a mismatch in position and momentum does not cause a loss in contrast of the interferometer. Further, we show that constraining the dephasing rate leads to stricter bounds on the noise parameters than enforcing a contrast threshold, indicating that good dephasing control ensures high interferometric contrast.
}
\end{abstract}
\maketitle


\section{Introduction}

Matter-wave interferometers have a multitude of applications ranging from quantum sensors to test the fundamental physics in a laboratory, see for e.g. ~\cite{Doherty_2013,Canuel_2018}. One particular material much sought for creating such matter-wave interferometers, which stands out, is the nanodiamond with a color-center defect, such as a nitrogen-vacancy (NV)-center. These nanoparticles have applications ranging from quantum metrology to quantum sensors due to the large coherence time scale of the NV spin even at room temperature~\cite{Bar_Gill_2013,Doherty_2013,Abobeih_2018,Wood_2022}. Combining the advancements in quantum technology and quantum material, it opens up a new vista to test fundamental physics with the help of NV-centred nanodiamond, especially for testing the quantum nature of spacetime in a lab, see ~\cite{Bose:2017nin,Marshman:2019sne,Bose:2022uxe}, and also
~\cite{Marletto:2017kzi}. These papers utilized the spin entanglement witness \cite{Bose:2017nin} to test the quantum nature of gravity~\cite{Gupta}, for a review see~\cite{michael1979advanced,Donoghue:2017pgk}. If two masses in quantum superpositions
can be entangled solely via gravity, then the spacetime
ought to behave like a quantum entity, see \cite{Marshman:2019sne,Bose:2022uxe,Vinckers:2023grv}, see also~\cite{Carney_2019,Danielson:2021egj,Christodoulou:2022mkf}, known as
the QGEM (Quantum Gravity Mediated Entanglement
of Masses) protocol. Entanglement between two masses also tests post-Newtonian corrections to low-energy quantum gravity~\cite{Toros:2024ozu}, relativistic corrections to the Coulomb potential \cite{Toros:2024ozf}, massive graviton in the context of
brane-world scenarios \cite{elahi2023probing}, $f(R)$ theories of quantum gravity, see~\cite{Vinckers:2023grv,chakraborty2023distinguishing}, axion physics \cite{Barker:2022mdz}, test of quantum version of the equivalence principle \cite{Bose:2022czr}, and 
entanglement between matter and photon degrees of
freedom \cite{Biswas:2022qto}, which will provide a quantum 
entanglement version of the light-bending experiment due
to the quantum natured massless spin-2 graviton~\cite{Gupta,michael1979advanced,Donoghue:2017pgk}.

The success of these important experiments crucially hinges upon maintaining the coherence and controlling the decoherence due to many external interactions, which are random~\cite{bassireview}. In general, any matter-wave interferometer is sensitive to external noise and fluctuations in ambient pressure, temperature, current, voltage, spin, electromagnetic field, etc.
\cite{Toros:2020dbf,Rijavec:2020qxd,
vandeKamp:2020rqh,Schut:2021svd,Schut:2023eux,Schut:2023hsy,Fragolino:2023agd,Schut:2024lgp,Schut:2023tce,Zhou:2025jki}. There are phonon-induced noise \cite{Henkel:2021wmj,Henkel:2023tqe,Xiang:2024zol},
and fluctuation in the spin degrees of freedom during the
dynamics of rotation of the rigid body \cite{Japha:2022phw,Zhou:2024pdl,Rizaldy:2024viw}, all leading
to dephasing and decoherence, and the loss of contrast \cite{Englert1988,Schwinger1988,Scully,Margalit:2020qcy}, infamoulsy known as the Humpty-Dumpty problem.

As we attempt to increase the size of the superposition, e.g., to test the QGEM protocol, inevitably, the noise, decoherence rate will also scale up, and it is paramount to understand how to control the decoherence rates. One dominant source of noise comes from the fluctuations in the electromagnetic  sector~\cite{Schut:2023tce,Fragolino:2023agd,HP_Noise}. The latter is very important, because the QGEM protocol will utilise a diamagnetic levitation~\cite{Bose:2017nin,DUrso16_GM,elahi2024} via fixed magnet or current-carrying wires. Then, a Stern-Gerlach mechanism is initiated by the magnetic field gradients, any fluctuations in the current will induce the fluctuations in the magnetic field and gradients, which form the core of such noise. How much of such a noise can be tolerated, then becomes an important question intrinsic to the process of wavefunction splitting and recombination at the end of one-loop interferometer~\cite{Margalit:2020qcy,Folman2013,folman2019,Folman2018}.

In this work, we will investigate a particular scheme for generating large spatial superpositions, on the order of $\sim 1\,\mu\mathrm{m}$, for a spin-1 NV center embedded in a nanodiamond, with mass $\sim 10^{-15}\,\mathrm{kg}$, within a time scale shorter than one second~\footnote{There are now many important schemes for creating macroscopic spatial superposition in a diamagnetic trap via the Stern Gerlach mechanism, see~\cite{scala2013matter,Pedernales:2020nmf,PhysRevLett.117.143003,steiner2024pentacene,Marshman:2021wyk,Zhou:2022epb,Zhou:2022jug,Zhou:2022frl}.}. Further, \cite{Pino_2018} proposes the use of inverted harmonic oscillator potential to accelerate the formation of large superpositions for mesoscopic objects.

A protocol which utilised the inverted harmonic oscillator potential~\cite{PhysRevA.111.052207} in a Stern-Gerlach type interferometer consists of five sequential stages. A very similar proposal has also been advertised in~\cite{Braccini:2024fey}, however, we will be following closely the sequences of \cite{PhysRevA.111.052207}.
Nevertheless, we will slightly depart crucially from \cite{PhysRevA.111.052207} to enhance the spatial superposition by nearly one order of magnitude. We will discuss the difference in the paper and in the appendix (by performing a comparison). Our approach departs from Ref.~\cite{PhysRevA.111.052207} in the timing of the initial stage: we choose its duration such that each state attains {\it maximum velocity at the end of the stage}, rather than {\it maximum spatial separation}, assumed in \cite{PhysRevA.111.052207}.  

Based on our renewed setup, we will further perturb the Lagrangian (by perturbing the external magnetic field, which helps to trap and create the superposition), and analyse the system’s susceptibility to external noise by computing the transfer function during the inverted-harmonic-potential stage. Using previously reported results for the magnetic noise in a harmonic-potential stage~\cite{HP_Noise}, we will determine the bounds on noise parameters under the assumption of white-noise statistics, ensuring that the dephasing rate ($\Gamma$) gives a coherence($e^{-\Gamma\tau}$, where $\tau$ is the total experimental time) of at least $10\%$ at the end of the interferometer ($e^{-\Gamma\tau}\geq0.1$). The same methodology can be extended to other noise models to obtain corresponding parameter constraints. However, for simplicity, we will stick to the case of white noise.

Benchmarking against contrast-based limits ($\mathcal{C}\geq0.9$), we find that constraining the dephasing rate imposes much tighter bounds on magnetic-field curvature, while bounds on the gradient remain comparable. Thus, in our scheme, a dephasing constraint is sufficient to ensure high interferometric contrast.

    
\section{Setup of the diamagnetic system}
We consider a spin-$1$ quantum system, such as a nitrogen-vacancy (NV) centre embedded in a diamond nanoparticle, with a relatively large mass of approximately \( m \sim 10^{-15} \, \mathrm{kg} \). The spin state of the particle is prepared in an equal superposition of the \( +1 \) and \( -1 \) eigenstates of the \( S_x \) operator. Our objective is to spatially separate these spin components, hence the centre of mass of the nanodiamond, to create a large spatial superposition on the order of \(\Delta x \sim 1 \, \mu \mathrm{m} \). We shall focus only on the dynamics along the x-axis (one-dimensional matter-wave interferometer), along which spatial superposition is intended to be created. The corresponding Lagrangian of the system is:
\begin{equation}
    L = \frac{1}{2}m\dot{x}^2 +\frac{\chi_\rho m}{2\mu_0}B_x^2 - \hbar\gamma_e S_x B_x - \hbar DS_{NV}^2 \label{eq.EffPSD_genLag}
\end{equation}
The parameter \( D \) denotes the zero-field splitting, which for NV centers is \( D = 2.87 \, \mathrm{GHz} \) \cite{Gruber2012-1997}.
The first term is the kinetic term for the nanodiamond, the second term is the diamagnetic induced contribution, where $\chi_{\rho}= -6.286\times10^{-9} \text{ m}^3 \text{ kg}^{-1}$ (for nanodiamond) represents the mass magnetic susceptibility, ~$\mu_0=4\pi\times 10^{-7}\text{ H m}^{-1}$ is the magnetic permeability, $\hbar=1.05\times 10^{-34}\text{ kg m}^2\text{ s}^{-1}$ is the reduced Planck's constant, $\gamma_e= 1.761 \times 10^{11}\text{ s}^{-1} \, \text{ T}^{-1}$ is the electron gyromagnetic ratio. The external magnetic field is given by $B_x$.  
We will use this Lagrangian to determine how to enhance the spatial superposition, $\Delta x$, and then we will study how susceptible the dephasing rate of the matter-wave interferometer due to fluctuations in the external magnetic field. Let us briefly recap the noise analysis in the following section. 

\section{General Framework to Compute Transfer Function and Dephasing}\label{sec.TFandD}
We outline the general framework to estimate the impact of external noise on interferometric phase evolution by computing effective transfer functions and the associated dephasing following Ref~\cite{HP_Noise}. We enumerate them below:
\begin{itemize}
    \item We begin with a general Lagrangian describing the motion in each arm \( j = R, L \) of the interferometer:
    \begin{align}
        L_j = \frac{1}{2}mv_j^2 - A_jx_j^2 - B_jx_j - C_j
    \end{align}
    where \( A_j, B_j, C_j \) are parameters characterising the system's potentials, which acquire time-dependent fluctuations due to experimental noise.

    \item The accumulated interferometric phase difference between the two paths is given by:
    \begin{align}
        \Delta \phi = \frac{1}{\hbar}\int_{t_i}^{t_f} (L_R - L_L)\,dt
    \end{align}
    In a noiseless setup, this phase depends purely on the deterministic difference in Lagrangians. However, fluctuations \( \delta A_j, \delta B_j, \delta C_j \) introduce stochastic phase contributions \( \delta\phi \), leading to dephasing.

    \item Assuming fluctuations affect only the Lagrangian coefficients (and not the trajectories), the stochastic phase deviation becomes:
    \begin{align}
    \delta \phi = &\frac{1}{\hbar}\int_{t_i}^{t_f} \bigg[(\delta A_Lx_L^2-\delta A_Rx_R^2) \nonumber\\
    &+ (\delta B_L x_L-\delta B_R x_R) + (\delta C_L-\delta C_R)\bigg]\,dt \label{eq.genphasedif}
    \end{align}
    The same can be extended to include fluctuations in the trajectories as well. In our analysis, we do not have fluctuations in terms of the form $C_j$.

    \item Since noise is inherently stochastic, we analyse its effect statistically via the ensemble-averaged phase variance:
    \begin{align}
        \Gamma = \lim_{\tau\to\infty} \frac{1}{\tau} E[\abs{\delta\phi(\tau)}^2] \label{eq.Gamma_form}
    \end{align}
    For simplicity, let's assume that the fluctuations in each of the coefficients ($\delta A_j$, $\delta B_j$ and $\delta C_j$) arise from the same noise source, $\delta\xi$. For example, this noise could arise from fluctuations in $B_x$ in Eq.~\eqref{eq.EffPSD_genLag}. The power spectral density $S(\omega)$ of the noise source is defined using the statistics of the fluctuations in $\xi$:
    \begin{equation}
        E[{\delta\widetilde{\xi}}(\omega){\delta\widetilde{\xi}^*}(\omega')] = S_{\xi\xi}(\omega)\delta(\omega-\omega')
    \end{equation}
    Using the Wiener-Khinchin theorem~\cite{Khintchine1934,Wiener:1930}, we can is recast Eq.~\eqref{eq.Gamma_form} in the frequency space as:
    \begin{align}
        \Gamma &= \int_{\omega_{\min}}^{\infty} S(\omega)\left|F(\omega)\right|^2\,d\omega\nonumber\\
        &\leq \int_{\omega_{\min}}^{\infty} S(\omega)\left|\sum_i \sqrt{F_i(\omega)}\right|^2\,d\omega \label{eq.Gamma_def}
    \end{align}
    where:
    \begin{itemize}
        \item \( S(\omega) \): power spectral density (PSD) of the noise,
        \item \( F(\omega) \): total effective system-dependent transfer function,
        \item \( F_i(\omega) \): transfer functions each accounting for different features of the system,
        \item \( \omega_{\min} = 2\pi / T_{\text{exp}} \): lower cutoff-frequency based on experiment duration ($T_{\text{exp}}$); frequencies much below $T_{\text{exp}}$ will behave like a static offset.
    \end{itemize}

    \item The transfer functions \( F_i(\omega) \) quantify the system's susceptibility to noise at different frequencies. Assuming spin-symmetric noise couplings (e.g., \( \delta A_L = \delta A_R \)), we define:
    \begin{align*}
        F_1(\omega) &\propto \left| \int_{t_i}^{t_f} (x_R - x_L) e^{i \omega t} dt \right|^2 \\
        F_2(\omega) &\propto \left| \int_{t_i}^{t_f} (x_R^2 - x_L^2) e^{i \omega t} dt \right|^2
    \end{align*}
    For spin-antisymmetric noise coupling (e.g., \( \delta B_L = -\delta B_R \)), the following also contribute:
    \begin{align*}
        F_3(\omega) &\propto \left| \int_{t_i}^{t_f} (x_R + x_L) e^{i \omega t} dt \right|^2 \\
        F_4(\omega) &\propto \left| \int_{t_i}^{t_f} (x_R^2 + x_L^2) e^{i \omega t} dt \right|^2
    \end{align*}
\end{itemize}

This framework enables a direct mapping between physical noise sources and their dephasing impact via the frequency-dependent response encoded in \( F_i(\omega) \) and the spectral properties of the noise encoded in \( S(\omega) \).

\section{Harmonic Potential – One-stage Process for creating spatial superposition} \label{sec.3A}
In the presence of a linear magnetic field gradient, the \( +1 \) and \( -1 \) spin states experience harmonic potentials that are displaced in opposite directions due to the difference in their magnetic moments~\cite{Pedernales:2020nmf,Marshman:2021wyk}, similar to the Stern Gerlach apparatus, for a review see~\cite{Folman2013,Folman2018,folman2019}. Consequently, each component of the spin superposition evolves under a distinct harmonic potential. If the particle is initially localised at the zero of the magnetic field gradient, the most probable positions of the two spin components will evolve in opposite directions, effectively producing a spin-centre-of-mass position entangled state. 

In this setup, the external magnetic field is taken to be~\footnote{We assume that the magnetic field is in the $x-y$ plane, therefore, $B_y=-\eta_0 y \hat y$. However, we are assuming that the superposition will take place in one dimension. We are assuming an ideal case where we take the initial condition of $y=0$. In reality, it will be extremely hard, and this will require knowing the centre-of-mass motion along $x,z$ directions extremely well. We will need to initiate the experiment at $y=0$, in which case there will be no displacement due to the external inhomogeneous magnetic field along this direction. 
}
\begin{equation}
\mathbf{B} = (B_0 + \eta_0 x)\hat{x}.
\end{equation} Hence the Lagrangian for each spin component \( j \in \{R, L\} \) is given by:
\begin{align}
    L_j = & \frac{1}{2} m v_j^2 - \frac{1}{2} m \omega_0^2 x_j^2 
    - \Big(S_{xj}\hbar \gamma_e \eta_0 - \frac{\chi_\rho m}{\mu_0} B_0 \eta_0 \Big)x_j \nonumber \\
    & + \frac{\chi_\rho m}{2 \mu_0} B_0^2 - S_{xj}\hbar \gamma_e B_0 - \hbar D S_{NV}^2, \label{eq.Lag_j_HP}
\end{align}
where the indices \( R \) and \( L \) correspond to the spin states \( S_{xR} = +1 \) and \( S_{xL} = -1 \), which evolve into the right and left arms of the interferometer, respectively. Here, \( \chi_{\rho} = -6.286 \times 10^{-9} \, \mathrm{m}^3 \, \mathrm{kg}^{-1} \) (for nanodiamond) represents the mass magnetic susceptibility of the particle. The characteristic frequency \( \omega_0 \) is defined as \cite{Pedernales:2020nmf, Marshman:2021wyk}:
\begin{align}
    \omega_0 = \left( -\frac{\chi_\rho}{\mu_0} \right)^{1/2} \eta_0. \label{eq.omega0_exp}
\end{align}
The maximum spatial separation between the two spin components that can be created using this potential alone is:
\begin{align}
    \Delta x_{\text{max}} &= \left| x_R(t) - x_L(t) \right|_{t = \frac{\pi}{\omega_0}}  \nonumber \\
    &= \left| \frac{4 \hbar \gamma_e \eta_0}{m \omega_0^2} \right|. \label{eq.max_sep}
\end{align}
As an illustration, to achieve a superposition size of \( \Delta x\sim  1 \, \mu \mathrm{m} \), the required magnetic field gradient must not exceed \( \eta_0 \approx 15 \, \mathrm{T \, m^{-1}} \). This corresponds to a time period \( T = \frac{2 \pi}{\omega_0} \approx 6 \, \mathrm{s}. \). This means we will have to keep the spin coherence that long to achieve a decent spatial superposition.

During the creation of the spatial superposition, the spin states are susceptible to decoherence, see~\cite{Zhou:2025jki}. Therefore, it is crucial to generate the desired separation while minimising decoherence to enable a spin interference measurement at the output of the interferometer. Additionally, it has been found~\cite{HP_Noise} that low magnetic field gradients can result in higher dephasing of the system due to magnetic noise. This motivates the exploration of alternative potential schemes capable of generating large spatial superpositions on shorter timescales. 


\section{New Inverted Harmonic Potential - Five stage Process}\label{sec.IHP-5stage}

Following the approach outlined in Ref.~\cite{PhysRevA.111.052207}, we consider a five-stage protocol designed to generate large spatial superpositions in a spin-1 system. However, here we will deviate from the original description of \cite{PhysRevA.111.052207}, and come up with a slightly different plan, which enables us to enhance the superposition by roughly one order of magnitude.

The initial state is prepared as an equal superposition of the $\ket{+1_x}$ and $\ket{-1_x}$ eigenstates of the spin operator along the $x$-axis. The subsequent stages are as follows:
\begin{enumerate}
    \item \textbf{Initial separation stage due to harmonic potential}: In this stage, we will use the harmonic potential generated through a linear magnetic field gradient, as described in the previous section, to entangle the spin with the position of the particle. A small bias magnetic field ($\sim 1\,\mathrm{mT}$) is sufficient to define the spin quantisation axis; see Appendix~\ref{Appendix A} for further discussion. Note that the bias field is not necessary to create the force or spatial superposition, but to avoid Majorana Spin flip~\cite{Brink_2006}. In contrast to the approach in Ref.~\cite{PhysRevA.111.052207}, where the evolution during this stage concludes at the point of maximum spatial separation and zero velocity, the present scheme is designed to terminate this stage at the point of maximum attainable velocity, given the initial conditions. This choice enables a more rapid subsequent evolution under the inverted potential. The duration of this stage is expected to be around 5ms.
    
    \item \textbf{Enhancement stage due to inverted harmonic potential}: To accelerate the formation of the spatial superposition, the second stage employs an inverted harmonic potential, realized through a magnetic field configuration with negative curvature\footnote{Magnetic field curvature refers to the second spatial derivative of the magnetic field; In our case it is $\frac{\partial^2 B}{\partial x^2}$\label{footnote.curvature}}. In this scheme, the wavepackets enter this stage with non-zero velocity, resulting in a faster and more efficient expansion of the superposition compared to the previous proposal~\cite{PhysRevA.111.052207}. Note that the inverted harmonic potential can be independent of the spin, unlike the previous stage. This is because a spin-dependent potential is needed only to initiate the creation of a spatial superposition. Consequently, the spin degree of freedom can be effectively ``switched off''. In practice, this can be achieved by coherently mapping the electronic spin states to long-lived nuclear spin states via a $\pi/2$ microwave pulse~\footnote{The nucleus has a lower magnetic moment as compared to electrons, and the potential is effectively spin-independent.}, see~\cite{Abobeih_2018}. The duration of this stage is expected to be around $\sim 150$ ms.
    
    \item \textbf{Return stage in harmonic potential}: To close the interferometer geometry, the velocities of the spatially separated wavepackets must be reversed. This is achieved by switching on a linear magnetic field gradient - a harmonic potential. The purpose of this stage is only to reverse the direction of the spatial evolution, and hence, the time taken in this stage can be minimised. In this stage, the maximum superposition size is achieved. For symmetry, we choose the return stage to be such that the velocity of each state at the end of the stage is exactly the opposite of that at the beginning of the stage. The duration of this stage is expected to be around $\sim 0.5$ ms, which is the shortest stage.
    \item \textbf{Deceleration stage in inverted harmonic potential}: We now turn on the IHP to decelerate the wavepackets. For symmetry, we keep the duration of this stage equal to the second stage. The duration of this stage is expected to be around $\sim 150$ ms.
    \item \textbf{Final stage of recombination in a harmonic potential}: To recombine the interferometer arms with minimal residual velocity, the spin degree of freedom is reactivated by mapping the nuclear spin states back to the electronic spin states using a $\pi/2$ microwave pulse. A spin-dependent harmonic potential, similar to the initial stage, is then applied to close the trajectories with zero velocity~\footnote{For a purely spin-independent harmonic potential acting on spatially separated trajectories, exact recombination at zero velocity is not achievable.}. The duration of this stage is expected to be around $\sim 5$ ms.
\end{enumerate}

Now we will briefly describe the equations of motion and the expected time taken by the one-loop interferometer. Along with the equations, we will also mention some values of the parameters that can produce a one micron superposition.

\subsection{Equations of motion: Initial Separation stage in harmonic potential}
In the initial separation stage, the external magnetic field is taken to have the form
\begin{equation}
    B(x) = B_{0H} + \eta_1 x , \label{eq.B_HP_form}
\end{equation}
where \(B_{0H} = 1\)mT and \(\eta_1 = 5\times 10^{3}\)T/m denotes the linear field gradient. In the simulation, Fig.\ref{fig:B_0-1_trajectory complete approx1micron_latestscheme_nomodellabel} all harmonic potential generating gradients are taken to be equal: $\eta_1 = \eta_3 =\eta_5$ (where the subscript 'n' denotes the parameter in the n-th stage of the protocol).
The Lagrangian for each spin component is analogous to Eq.~\eqref{eq.Lag_j_HP}, but with the characteristic frequency \(\omega_0\) replaced by
\begin{equation}
\omega_1 = \bigg(-\frac{\chi_\rho}{\mu_0}\bigg)^{1/2} \, \eta_1 ,
\end{equation}
where the subscript “1” labels the first stage of the protocol.
The equation of motion is given by:
\begin{align}
    m\ddot{x}_j(t)&= -m\omega_1^2x_j(t) - C_j\eta_1\label{eq.EOM_det}
\end{align}
where, 
\begin{align}
    C_j = \big( S_{xj}\hbar\gamma_e - \frac{ \chi_\rho m}{\mu_0}B_{0H}\big)
\end{align}
Imposing $x_j(0) = 0$ and $\dot{x}(0) = 0$, we get
\begin{align}
    x_j(t) = \frac{C_j \eta_1}{m\omega_1^2 } (\cos(\omega_1 t) - 1) \label{eq.det_traj}
\end{align} 
In Ref.~\cite{PhysRevA.111.052207}, the duration of this stage was chosen so that the particle reached its maximum spatial separation within the given potential, which occurs when the velocity vanishes.  
In contrast, in the present scheme we terminate the stage when each component attains its \emph{maximum velocity}.  
This choice ensures that, upon switching to the inverted harmonic potential in stage~2, the spatial separation grows more rapidly, thereby reducing the total time required to achieve a target superposition size compared to the method of Ref.~\cite{PhysRevA.111.052207} (For further comparison, refer to Appendix~\ref{Appendix B}.

Let $T_1$ be the duration of stage 1. At $T_1 = \frac{\pi}{2\omega_1}$, 
\begin{align}
    x_j(T_1) &= \frac{-C_j \eta_1}{m\omega_1^2} \equiv -N_{1j}\label{eq.x1}\\
    \dot{x}_j(T_1) &= \frac{-C_j \eta_1}{m\omega_1} \equiv -N_{1j}\omega_1 \label{eq.dotx1}
\end{align}
The parameter \(N_{1j}\) thus encapsulates the displacement amplitude for the \(j\)-th arm at the end of the initial separation stage (recall that $j\in{R,L}$ representing the right and left arms of the interferometer). 

\subsection{Equations of Motion: Enhancement stage in inverted harmonic potential}

\begin{figure*}[ht!]
    \includegraphics[width=\linewidth]{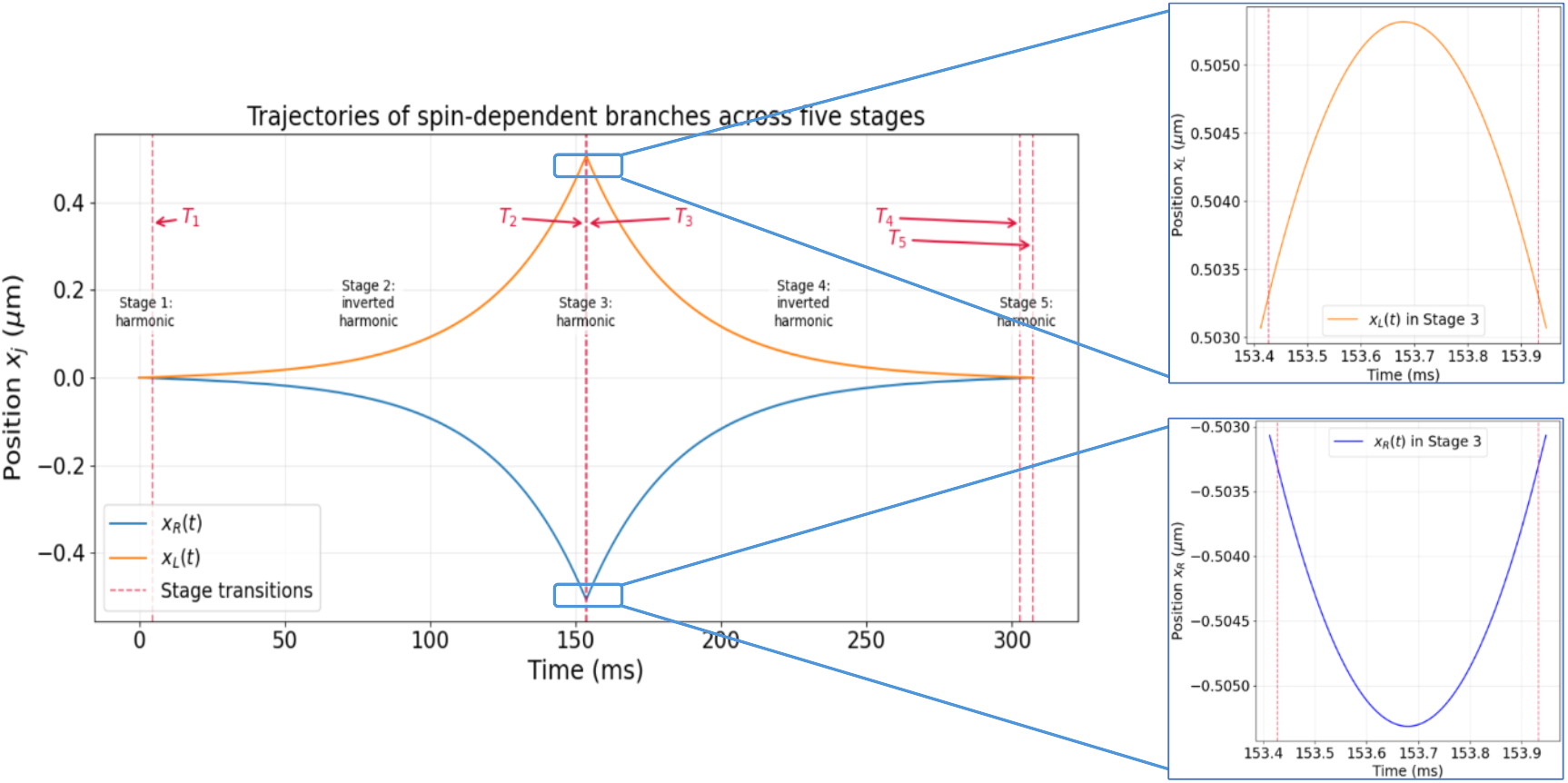}
    \captionsetup{justification=RaggedRight}
    \caption{Spin-dependent trajectories in a five-stage interferometric protocol for a spin-1 particle. This schematic represents the trajectories of the two spin states composing the superposition, in their center of mass frame. The trajectories have been obtained by plotting the EOM obtained in Sec.\ref{sec.IHP-5stage} and using the parameter values mentioned in Table.\ref{tab:stage_params}.
    The two arms (blue and orange) correspond to \( S_x = \pm 1 \) states. 
    Stage 1 (spin-dependent harmonic potential) generates spin-position entanglement and ends at \emph{maximum velocity}, in contrast with the previous scheme ~\cite{PhysRevA.111.052207} that stopped at maximum spatial separation. 
    This change enables faster and greater separation in Stage 2 (spin-independent inverted harmonic potential);
    Stage 3 (harmonic potential) reverses motion at peak separation; 
    Stage 4 (inverted harmonic potential) decelerates the arms; 
    and Stage 5 (spin-dependent harmonic potential) recombines them. Red dashed lines indicate stage boundaries. The zoom-in confirms a smooth trajectory even in Stage 3. 
    Field strengths remain below niobium's lower critical field $H_{c1}=135$ mT \cite{elahi2024,alekseevskiy2025,Hudson1971}, minimising flux noise in the Nb superconductors used.
    }
    \label{fig:B_0-1_trajectory complete approx1micron_latestscheme_nomodellabel}
\end{figure*}

In the enhancement stage, the external magnetic field is taken to be
\begin{equation}
B(x) = B_{0I} - \eta_2 x^2 ,\label{eq.B_IHP_form}
\end{equation}
where \(B_{0I} = 0.1\)T~\footnote{The value of the bias magnetic fields as well as the magnetic field gradients and curvatures are constrained to not give rise to magnetic fields beyond the lower critical value for type II superconductors, which ensures the minimisation of the flux noise in the superconductor. In case of Nb type II superconductors, the lower critical field, $H_{c_1} \sim 170$mT, and we choose all fields $B_0\lesssim 100$mT~\cite{Hudson1971, elahi2024}.} is a uniform bias field and \(\eta_2 = 1\times10^{6}\) Tm$^{-2}$ characterises the magnetic field curvature. In the simulation, Fig.\ref{fig:B_0-1_trajectory complete approx1micron_latestscheme_nomodellabel} all inverted harmonic potential generating curvatures are taken to be equal: $\eta_2 = \eta_4$. In this stage, the electronic spin has been mapped to the nuclear spin, and thus $S_{xj} = 0$ for both interferometer arms, as discussed earlier. Thus, the corresponding Lagrangian for each interferometer arm (j = R, L) is
\begin{align}
    L_j &= \frac{1}{2} m \dot{x}_j^2 
    + \frac{\chi_\rho m}{2\mu_0} B_{0I}^2
    + \frac{\chi_\rho m}{2\mu_0} \eta_2^2 x_j^4
    + \frac{1}{2} m \omega_2^2 x_j^2 ,\label{eq.IHPLag}
\end{align}
where
\begin{align}
    \omega_2^2 = -\frac{2\chi_\rho}{\mu_0} \eta_2 B_{0I} .
\end{align}
The subscript $2$ in $\omega_2$ reflects the frequency of the second stage. 
The corresponding equation of motion is
\begin{align}
    m\ddot{x}_j(t) &= m\omega_2^2 x_j(t) + \frac{2\chi_\rho m}{\mu_0} \eta_2^2 x_j^3(t) .\label{eq.EOMstage2_exact}
\end{align}
We neglect the cubic term in the dynamics, which is equivalent to omitting the quartic term in the Lagrangian. The linearised equation of motion then has the solution
\begin{align}
    x_j(t) &= x_j(T_1)\cosh\!\left[ \omega_2 t - \phi_2 \right] 
    + \frac{\dot{x}_j(T_1)}{\omega_2} \sinh\!\left[ \omega_2 t - \phi_2 \right] , \label{eq.x2traj}
\end{align}
where $\phi_2 = \omega_2 T_1$ ensuring continuity with the end of stage~1.

Let \(T_2\) denote the time at the end of stage~2~\footnote{This notation differs from Ref.~\cite{PhysRevA.111.052207}, where \(T_2\) refers to the \emph{duration} of the second stage. Here, the time variable is continuous from \(t = 0\) to \(t = T_5\) for the complete interferometer sequence.}.  
For mathematical convenience, we choose the duration of this stage as $$T_2 - T_1 = \frac{3\pi}{2\omega_2}.$$ 
Durations in the vicinity of this value produce superposition sizes of the same order of magnitude. The choice reflects a trade-off between minimising decoherence (which grows with the time interval) and maximising spatial separation (which also increases with the duration of the inverted harmonic evolution).

Using Eqs.~\eqref{eq.x1}-~\eqref{eq.dotx1} and Eq.~\eqref{eq.x2traj}, the position and velocity at t \(t = T_2\) are given by:
\begin{align}
    x_j(T_2) &= -N_{1j} \left[ \cosh(1.5\pi) + \frac{\omega_1}{\omega_2} \sinh(1.5\pi) \right] , \label{eq.x2} \\
    \dot{x}_j(T_2) &= -N_{1j} \left[ \omega_2 \sinh(1.5\pi) + \omega_1 \cosh(1.5\pi) \right] . \label{eq.dotx2}
\end{align}

Since \(\cosh(1.5\pi) \approx \sinh(1.5\pi) \approx e^{1.5\pi}/2\) to three significant figures,  
Eqs.~\eqref{eq.x2}–\eqref{eq.dotx2} simplify to
\begin{align}
    x_j(T_2) &\approx -N_{1j} \frac{e^{1.5\pi}}{2} \left( 1 + \frac{\omega_1}{\omega_2} \right) , \label{eq.x2approx} \\
    \dot{x}_j(T_2) &\approx -N_{1j} \frac{e^{1.5\pi}}{2} \left( \omega_2 + \omega_1 \right) . \label{eq.dotx2approx}
\end{align}
These expressions will be used as the initial conditions for the return stage.

\subsection{Equations of Motion: Return stage in a harmonic potential}

In the return stage, the external magnetic field is similar to Eq.~\eqref{eq.B_HP_form}, with $\eta_1$ replaced with \(\eta_3\) denotes the linear gradient. 
The Lagrangian in this stage has the same form as Eq.~\eqref{eq.Lag_j_HP}, but with \(S_{xj} = 0\) since the electronic spin remains mapped to the nuclear spin.  
The corresponding harmonic frequency in the third stage is given by:
\begin{align}
    \omega_3 = -\bigg(\frac{\chi_\rho}{\mu_0}\bigg)^{1/2} \, \eta_3 .
\end{align}

The general solution to the linearised equation of motion is
\begin{align}
    x_j(t) &= A_{3j} \cos(\omega_3 t) + B_{3j} \sin(\omega_3 t) - \frac{B_{0H}}{\eta_3} .
\end{align}
Applying the continuity conditions at \(t = T_2\), given by Eqs.~\eqref{eq.x2} and \eqref{eq.dotx2}, the solution may be expressed as
\begin{align}
    x_j(t) &= \left( x_j(T_2) + \frac{B_{0H}}{\eta_3} \right) \cos(\omega_3 t - \phi_3) \nonumber\\
    &\quad + \frac{\dot{x}_j(T_2)}{\omega_3} \sin(\omega_3 t - \phi_3) - \frac{B_{0H}}{\eta_3} ,
\end{align}
where $\phi_3 = \omega_3 T_2$. This can be rewritten in the compact form
\begin{align}
    x_j(t) &= N_{3j} \, \sin\!\left( \omega_3 t - \phi_3 + \widetilde{\phi}_{3j} \right) - \frac{B_{0H}}{\eta_3} ,
    \label{eq.argument_for_superposition}
\end{align}
with
\begin{align}
    N_{3j} &= \sqrt{\left( x_j(T_2) + \frac{B_{0H}}{\eta_3} \right)^2 + \frac{\dot{x}_j^2(T_2)}{\omega_3^2}} , \\
    \widetilde{\phi}_{3j} &= \arctan\!\left[ \frac{\left( x_j(T_2) + \frac{B_{0H}}{\eta_3} \right) \omega_3}{\dot{x}_j(T_2)} \right] \pm n\pi .
\end{align}
The maximum spatial separation in this stage occurs when the sine term in Eq.~\eqref{eq.argument_for_superposition} reaches \(\pm 1\), i.e., when its argument equals \(n\pi/2\) for the smallest positive integer \(n\). Let \(T^*\) denote the interval between \(T_2\) and the first such maximum spatial superposition that can be obtained. Since \(B_{0H}\) contributes only an offset and does not affect the separation magnitude, we may set \(B_{0H} = 0\) in the following the computation of the maximum superposition size~\footnote{Note that the $N_{3j}$ and $N_{1j}$ are each the same for both j=R and j=L, when $B_{0H}=0$}:
\begin{align}
    \Delta x_{\mathrm{max}} &= 2 N_{3j} \big|_{B_{0H} = 0} \quad \text{(for either $j = R, L$)} \nonumber\\
    &= 2 N_{1j} \big|_{B_{0H} = 0} \, \frac{e^{1.5\pi}}{2}
    \sqrt{\left( 1 + \frac{\omega_1}{\omega_2} \right)^2 + \left( \frac{\omega_1 + \omega_2}{\omega_3} \right)^2} \nonumber\\
    &= \frac{\hbar \gamma_e \eta_1}{m \omega_1^2} \, e^{1.5\pi} (\omega_1 + \omega_2)
    \sqrt{\frac{1}{\omega_2^2} + \frac{1}{\omega_3^2}} .
\end{align}
where, $N_{1j}$ is defined in Eq.~\eqref{eq.x1}. The duration of the return stage is taken as $$T_3 - T_2 = 2 T^*,$$ ensuring that the trajectories complete the intended reversal of motion.  
At the end of this stage, the positions and velocities are
\begin{align}
    x_j(T_3) &= N_{3j} - \frac{B_{0H}}{\eta_3} , \label{eq.x3} \\
    \dot{x}_j(T_3) &= N_{3j} \, \omega_3 . \label{eq.dotx3}
\end{align}

\subsection{Equations of Motion: Deceleration stage in a harmonic potential}

In the deceleration stage,  the external magnetic field is similar to Eq.~\eqref{eq.B_IHP_form}, with $\eta_2$ replaced with \(\eta_4\) denotes the magnetic field curvature. The Lagrangian in this stage is analogous to Eq.~\eqref{eq.IHPLag}, and the associated characteristic frequency in the fourth stage is given by:
\begin{align}
    \omega_4^2 = -\frac{2\chi_\rho}{\mu_0} \, \eta_4 B_{0I} .
\end{align}

Neglecting the cubic term in the equation of motion, as in stage~2, the solution for the \(j\)-th interferometer arm is
\begin{align}
    x_j(t) = x_j(T_3) \cosh\!\left( \omega_4 t - \phi_4 \right) 
    + \frac{\dot{x}_j(T_3)}{\omega_4} \sinh\!\left( \omega_4 t - \phi_4 \right) ,
\end{align}
where $\phi_4 = \omega_4 T_3$. The initial conditions for this stage correspond to the final position and velocity of stage~3, but with the velocity direction reversed.  
If we set \(\omega_4 = \omega_2\), then by time-reversal symmetry the trajectory will be symmetric about the point of maximum superposition, provided the stage duration is
\begin{equation}
T_4 - T_3 = \frac{1.5\pi}{\omega_4} = \frac{1.5\pi}{\omega_2} .
\end{equation}
Under this choice, the position and velocity at the end of stage~4 are
\begin{align}
    x_j(T_4) &= x_j(T_3) \cosh(1.5\pi) 
    + \frac{\dot{x}_j(T_3)}{\omega_4} \sinh(1.5\pi) , \label{eq.x4} \\
    \dot{x}_j(T_4) &= x_j(T_3) \, \omega_4 \sinh(1.5\pi) 
    + \dot{x}_j(T_3) \cosh(1.5\pi) . \label{eq.dotx4}
\end{align}
These expressions serve as the initial conditions for the recombination stage.

\subsection{Equations of Motion: Recombination stage in a harmonic potential}

In the recombination stage,  the external magnetic field is similar to Eq.~\eqref{eq.B_HP_form}, with $\eta_1$ replaced with \(\eta_3\) denotes the linear gradient. The Lagrangian, equation of motion, and associated parameters are analogous to those in stage~1, but with the position and velocity initial conditions taken from the end of stage~4. Imposing Eqs.~\eqref{eq.x4} and \eqref{eq.dotx4} as the initial conditions at \(t = T_4\), the general solution is
\begin{align}
    x_j(t) &= x_j(T_4) \cos(\omega_5 t - \phi_5)
    \nonumber\\
    &\quad+ \frac{\dot{x}_j(T_4)}{\omega_5} \sin(\omega_5 t - \phi_5)
    - \frac{C_j \eta_5}{m \omega_5^2} ,
\end{align}
where $\phi_5 = \omega_5 T_4$. This can be rewritten in the compact form
\begin{align}
    x_j(t) &= N_{5j} \, \sin\!\left( \omega_5 t - \phi_5 + \widetilde{\phi}_5 \right)
    - \frac{C_j \eta_5}{m \omega_5^2} ,
\end{align}
with
\begin{align}
    N_{5j} &= \sqrt{ x_j^2(T_4) + \frac{\dot{x}_j^2(T_4)}{\omega_5^2} } , \\
    \widetilde{\phi}_5 &= \arctan\!\left( \frac{x_j(T_4) \, \omega_5}{\dot{x}_j(T_4)} \right) .
\end{align}

If \(\eta_5 = \eta_1\), then by symmetry with stage~1 the duration of this stage is chosen as
\begin{equation}
T_5 - T_4 = \frac{\pi}{2\omega_5} ,
\end{equation}
matching the separation time in stage~1.  
At \(t = T_5\), this ensures that the two interferometer arms recombine:
\begin{align}
    x_R(T_5) - x_L(T_5) &= 0 , \label{eq.x15} \\
    \dot{x}_j(T_5) &= 0 . \label{eq.dotx5}
\end{align}
\subsection{Some Remarks}
Note that a nonzero \(B_{0H}\) produces only a uniform offset in both trajectories and does not affect the recombination condition. However, we have neglected higher-order terms in the equations of motion during the inverted harmonic potential stages, as given in Eq.~\eqref{eq.EOMstage2_exact}. Including these terms may modify the recombination condition. 

In Fig.~\ref{fig:B_0-1_trajectory complete
approx1micron_latestscheme_nomodellabel}, the trajectories are obtained from
the above equations but are shown in the centre-of-mass frame, i.e., we take
\begin{align}
    &x_R(t) \to x_R(t) - x_{\mathrm{COM}}(t);\; x_L(t) \to x_L(t) - x_{\mathrm{COM}}(t)\nonumber\\
    &\text{where,} \quad x_{\mathrm{COM}}(t) = \frac{x_R(t) + x_L(t)}{2}
\end{align}
Equivalently, this corresponds to removing the uniform offset in the potential, which can be viewed as taking \(B_{0H}=0\). This choice is made purely for illustrative purposes, to highlight the relative motion of the two interferometer arms. In the noise analysis, however, we retain a non-zero magnetic field offset.

If we increase the duration of the third stage a little longer, such that the symmetry of the interferometer trajectory about the maximum superposition size is broken, the states would gain higher initial velocity when entering the inverted harmonic deceleration stage. Though this would help hasten the closing of the trajectories, it makes it harder to obtain a zero velocity at the close of the interferometer. Hence, the interferometer can become susceptible to the Humpty Dumpty problem. The same is true for errors that may creep into the experimental precision of ending a particular stage. We shall later analyse how much of the deviation in position and velocity at the closing of the interferometer can be tolerated.

\section{Noise Analysis}


The objective in this section is to evaluate the dephasing induced by fluctuations in both the gradient and the curvature of the magnetic field for the five-stage scheme described in Sec.~\ref{sec.IHP-5stage}.  
In Ref.~\cite{HP_Noise}, the dephasing was computed for the case of a purely harmonic potential.  
Here, we adapt those results to estimate the order of magnitude of noise effects in the harmonic-potential stages of the present protocol.

By symmetry of the interferometer sequence, the dephasing during stage~2 and stage~4 is identical.  
It therefore suffices to compute the dephasing for stage~2 and directly extrapolate the result to stage~4.

Our approach is as follows:  
First, we determine the effective transfer function of the inverted harmonic potential (IHP) stage, which characterises the susceptibility of the system to fluctuations at different noise frequencies.  
This transfer function encodes how efficiently magnetic-field noise at a given frequency contributes to dephasing.  
We then use it to quantify the effect of white noise in both the magnetic-field gradient and curvature.  
Finally, we derive upper bounds on the allowable noise amplitudes that keep the total dephasing within a chosen optimal limit.

For illustration, we will consider white noise, which is characterised by the following statistics~\footnote{We could consider Lorentzian noise profile or pink noise, but we realised in \cite{HP_Noise} that in general white noise causes more of a dephasing. Similar situations arise in noise induced directly in the spin of the NV, see~\cite{Zhou:2025jki}. This is because random fluctuations in both the arms of the matter-wave interferometer give rise to a larger uncorrelated noise, and hence lead to a larger dephasing. }:
    \begin{align}
        E[\delta\eta(t)] &= 0  \label{eq.whitenoise_statprop1}\\
        E[\delta\eta(t)\delta\eta(t')] &= A^2\delta(t - t')\label{eq.whitenoise_statprop2}
    \end{align}
where $\eta$ is the fluctuating quantity with white noise statistics. From eq.(\ref{eq.whitenoise_statprop2}), we obtain the following PSD for white noise {~\footnotetext{$\delta\widetilde{\eta}_\tau(\omega) = \frac{1}{2\pi}\int_{-\tau}^{\tau}{\delta\eta}(t)\ e^{i\omega t}\,dt$\label{fn_S}}.}
\begin{align}
    S_{\eta\eta}(\omega) &=  \lim_{\tau \to \infty}\frac{1}{\tau}E[\delta\widetilde{\eta}_\tau(\omega)\delta\widetilde{\eta}_\tau^*(\omega)]  =A^2 \label{eq.gen_white_noise} 
\end{align} 
We consider white noise, since the fluctuations in the current that give rise to the magnetic field gradient fluctuations were found to be of the same order of magnitude for flicker noise and white noise in the harmonic potential stage~\cite{HP_Noise}. We, hence, assume that the bound on the fluctuations of the current can be obtained by analysis of one type of noise, and the result can be extended to other types of noise, and the bound on the parameters of other types of noise can be derived. 

\subsection{Computing the Transfer Function of the inverted harmonic potential stage}\label{sec.6A}

We first note that the equation of motion obtained via approximation, where we have ignored terms of  $\propto x^4$ has been ignored in the Lagrangian (we are only considering the leading order contribution, which suffices to ignore the $\propto x^4$ contribution), and each of the states in the superposition has $S_{xj} = 0$. We will use the same notation as used to describe the five-stage scheme. The Lagrangian describing the IHP stage is:
\begin{align}
    L_j = \frac{1}{2}m{\dot x_j}^2 +\frac{\chi_\rho m}{2\mu_0}B_{0I}^2 - \frac{1}{2}m\omega_2^2  x_j^2
\end{align}

Consider noise in the magnetic field curvature: 
\begin{align}
    \eta_2 = \eta_{02} + \delta \eta_2 (t)
\end{align}
The power spectral density corresponding to white noise in the curvature of the magnetic field is denoted by:
\begin{align}
        S_{\eta_2\eta_2}(\omega) &=  \lim_{\tau \to \infty}\frac{1}{\tau}E[\delta\widetilde{\eta}_2(\omega)\delta\widetilde{\eta}_2^*(\omega)]  =A_2^2 \label{eq.eta_2_white_noise} 
    \end{align} 
By definition of the "natural frequency" of the system, we have~\footnote{Note that $\delta \omega^2_2 (t)$ is not the same as $(\delta \omega_2)^2 (t)$.}:
\begin{align}
    \omega^2_2 &= \omega^2_{02} + \delta \omega^2_2 (t)\\
\text{where,}\quad 
    \delta \omega_2^2 &= \frac{-2\chi_\rho }{\mu_0}B_{0I} \delta \eta_2
\end{align}
This noise also causes fluctuations in the trajectory:
\begin{align}
    x_j^{tot} = x_j + \delta x_j
\end{align}

The equation of motion is:
\begin{align}
    m\ddot{x}^{tot}_j(t)&= m\omega_2^2x_j^{tot}(t)\nonumber\\
    m\ddot{x}_j(t) + m\delta \ddot{x}_j(t)&= m\omega_2^2x_j(t) + 2m\delta \omega^2_2 x_j(t) + m\omega_2^2\delta x_j(t)
\end{align}
where we have ignored higher order fluctuations and $x_j(t)$ satisfies the non-perturbed EOM Eq.~\eqref{eq.x2traj}. The perturbation in the trajectory has the following EOM:
\begin{align}
    \delta \ddot{x}_j(t)&= 2\bigg(x_j(T_1)\cosh(\omega_2 t -\phi_2) \nonumber\\
    &\quad\quad+ \frac{\dot{x}_j(T_1)}{\omega_2}\sinh(\omega_2 t -\phi_2)\bigg)\delta \omega^2_2 + \omega_2^2\delta x_j(t)\label{eq.deltax_eom_model2}
\end{align}
where all terms are as defined in Sec.\ref{sec.IHP-5stage}
Since we want to obtain the transfer function, which is a function of frequency, we carry out a Fourier transform~\footnote{where $\omega$ is the frequency in the Fourier space, which shall be physically understood as the frequency of the noise considered; $\omega$ is not to be confused with $\omega_2$.}.
\begin{align}
     \delta x_j(t)&=\int _{-\infty }^{\infty }{\delta \widetilde x_j}(\omega )\ e^{i\omega t}\,d\omega \label{eq.FTx} \\
     \delta\eta_2(t)&=\int _{-\infty }^{\infty }{\delta\widetilde{\eta}_2}(\omega )\ e^{i\omega t}\,d\omega \label{eq.FTeta}
\end{align} 
Imposing the EOM governing the perturbations in the trajectory, Eq.~\eqref{eq.deltax_eom_model2}, we obtain:
\begin{align}
    {\delta \widetilde x_j}(\omega)&= \frac{\alpha_j}{2(\omega_2^2+\omega^2)}\bigg(\cosh(\omega_2 t -\phi_2) \nonumber\\
    &\quad\quad+ \lambda\sinh(\omega_2 t -\phi_2)\bigg) {\delta\widetilde{\eta}_2}(\omega)
\end{align}
where $\alpha_{j}$ and $\lambda$ are defined as:
\begin{align}
    \alpha_{j} &= 8B_{0I}\frac{C_j}{m\eta_1},\label{eq.def_alpha_j}\\
    \lambda &= \frac{\eta_1^2}{2B_{0I}\eta_2}.\label{eq.def_lambda}
\end{align}
Now, we note down only that part of the difference in the Lagrangian contributed by fluctuations, and denote it by $\delta L$~\footnote{We will assume that the noise statistics affecting the left and the right arms are the same, else we have to account for the difference by considering different fluctuating quantities $\delta\eta_{2R}$ and $\delta\eta_{2L}$ explicitly.}:
\begin{align}
    \delta L =\ 
    &\underbrace{m(\dot x_R\delta \dot x_R - \dot x_L \delta \dot x_L)}_{\delta L^{(a)}} - \underbrace{m\omega_2^2 (x_R\delta x_R - x_L\delta x_L)}_{\delta L^{(b)}}\nonumber\\
    &\quad - \underbrace{\frac{1}{2}m\delta\omega_2^2 (x_R^2 - x_L^2)}_{\delta L^{(c)}}
    \label{eq.pertlag_2}
\end{align}
Let us discuss the above three contributions separately.
\begin{enumerate}
    \item 
The phase perturbation associated with \(\delta L^{(a)}\), denoted \(\delta\phi^{(a)}\), originates from velocity fluctuations, which in turn arise indirectly from variations in the magnetic-field curvature.  

\item
Similarly, \(\delta\phi^{(b)}\) corresponding to \(\delta L^{(a)}\) represents the phase perturbation resulting from trajectory fluctuations caused by the curvature variations.

\item
Finally, \(\delta\phi^{(c)}\) denotes the phase fluctuation induced directly by curvature fluctuations of the magnetic field.  

\end{enumerate}
Now we compute each of the phase fluctuations independently, starting with $\delta\phi^{(a)}$
\begin{align}
    &\delta\phi^{(a)} = \frac{1}{\hbar}\int_{T_1}^{T_2} m(\dot x_R\delta \dot x_R - \dot x_L \delta \dot x_L)\,dt\\
    &=- 16\beta\int_{T_1}^{T_2}\bigg(\frac{\omega_2}{\eta_1^2}\sinh(\omega_2 t -\phi_2) + \frac{\omega_1}{\eta_1^2}\cosh(\omega_2 t -\phi_2)\bigg)\nonumber\\
     &\quad\quad\times \bigg(\cosh(\omega_2 t -\phi_2) + \lambda\sinh(\omega_2 t -\phi_2)\bigg) \nonumber\\
     &\quad\quad\times\int_{-\infty}^{\infty} i\frac{\omega}{(\omega_2^2+\omega^2)}{\delta\widetilde{\eta}_2}(\omega) e^{i\omega t}\,d\omega \, dt \label{eq.delta_phia}
\end{align}
where,
\begin{equation}
    \beta = B_{0I}B_{0H}\gamma_e \label{eq.def_beta}
\end{equation}
Similarly, we compute the other phase fluctuations as well.
\begin{align}
    &\delta\phi^{(b)} = \frac{1}{\hbar}\int_{T_1}^{T_2} - m\omega_2^2 (x_R\delta x_R - x_L\delta x_L)\,dt\\
    &=32\beta\omega_2^2\int_{T_1}^{T_2}\int_{-\infty}^{\infty}d\omega  \, dt\,\bigg(\frac{1}{\eta_1^2}\cosh(\omega_2 t -\phi_2) \nonumber\\
    &\quad\quad+ \frac{1}{\eta_1\sqrt{2B_{0I}\eta_2}}\sinh(\omega_2 t -\phi_2)\bigg)\nonumber\\
     &\quad\quad\bigg(\cosh(\omega_2 t -\phi_2) + \lambda\sinh(\omega_2 t -\phi_2)\bigg)
     \frac{{\delta\widetilde{\eta}_2}(\omega)e^{i\omega t}}{2(\omega_2^2+\omega^2)} \label{eq.delta_phib}
\end{align}

\begin{align}
    &\delta\phi^{(c)} = \frac{1}{\hbar}\int_{T_1}^{T_2} - \frac{1}{2}m\delta\omega_2^2  (x_R^2 - x_L^2)\,dt\\
    &=4\beta\int _{-\infty }^{\infty }\int_{T_1}^{T_2}d\omega \, dt\,\bigg(\frac{1}{\eta_1}\cosh(\omega_2 t -\phi_2) \nonumber\\
    &\quad\quad+ \frac{1}{\sqrt{2B_{0I}\eta_2}}\sinh(\omega_2 t -\phi_2)\bigg)^2{\delta\widetilde{\eta}_2}(\omega ) e^{i\omega t} \label{eq.delta_phic}
\end{align}
First we note that, in general if $\delta\Phi= A+B$, and if we define $\delta\Phi_1= A$, $\delta\Phi_2= B$, then
\begin{align}
    \bigg(\sqrt{E[\delta\Phi_1\delta\Phi_1^*]}+\sqrt{E[\delta\Phi_2\delta\Phi_2^*]}\bigg)^2 \geq E[\delta\Phi\delta\Phi^*] \label{eq.ineq}
\end{align}
Hence, for mathematical ease, we shall compute each of the phase fluctuations independently and apply this identity at the end to obtain an upper bound on the maximum dephasing. First, we start with obtaining an effective transfer function corresponding to each of the phase fluctuation terms, $\delta\phi^{(a)}$, $\delta\phi^{(b)}$ and $\delta\phi^{(c)}$.

We carry out the ensemble average of the phase-variance using the EOM in the Fourier space, using Eq.~\eqref{eq.delta_phia}, Eq.~\eqref{eq.delta_phib} and Eq.~\eqref{eq.delta_phic}. The time-integrals~\footnote{The time integral runs from $T_1 = \pi/2\omega_1$ to $T_2 = 3\pi/2\omega_2$} involved in this step constitute the transfer function (as defined in Eq.~\eqref{eq.Gamma_def}. First, we compute the transfer function corresponding to the phase fluctuation $\delta\phi^{(a)}$:
\begin{widetext}
\begin{align}
    E[\delta\phi^{(a)}\delta\phi^{(a)*}] 
    &=(16\beta)^2\int_{T_1}^{T_2}\prod_{t = t_1,t_2}\bigg(\frac{\omega_2}{\eta_1^2}\sinh(\omega_2 t -\phi_2) + \frac{\omega_1}{\eta_1^2}\cosh(\omega_2 t -\phi_2)\bigg) \bigg(\cosh(\omega_2 t -\phi_2) + \lambda\sinh(\omega_2 t -\phi_2)\bigg) \nonumber\\
     &\quad\quad\quad\quad\times\int_{-\infty}^{\infty} \frac{\omega}{(\omega_2^2+\omega^2)}\frac{\omega'}{(\omega_2^2+\omega'^2)}E[{\delta\widetilde{\eta}_2}(\omega){\delta\widetilde{\eta}^*_2}(\omega')] e^{i\omega t_1-i\omega' t_2}\,d\omega\,d\omega'\,dt_1\,dt_2
\end{align}
From Eq.~\eqref{eq.eta_2_white_noise},
\begin{align}
    E[\delta\phi^{(a)}\delta\phi^{(a)*}] &=(16\beta)^2\int_{-\infty}^{\infty}\,d\omega \,S_{\eta_2\eta_2}(\omega)\frac{\omega^2}{(\omega_2^2+\omega^2)^2} \nonumber\\
     &\quad\quad\times\bigg|\int_{T_1}^{T_2}\bigg(\frac{\omega_2}{\eta_1^2}\sinh(\omega_2 t -\phi_2) + \frac{\omega_1}{\eta_1^2}\cosh(\omega_2 t -\phi_2)\bigg) \bigg(\cosh(\omega_2 t -\phi_2) + \lambda\sinh(\omega_2 t -\phi_2)\bigg)e^{i\omega t}\,dt\bigg|^2 \label{eq.TF_a_comp}
\end{align}
\end{widetext}
Using a definition of transfer function similar to that given in Eq.~\eqref{eq.Gamma_def};
\begin{equation}
    E[\delta\phi^{(a)}\delta\phi^{(a)*}] =\int_{-\infty}^{\infty}\,d\omega \,S_{\eta_2\eta_2}(\omega)F^{(a)}(\omega),
\end{equation}
and comparing it with Eq.~\eqref{eq.TF_a_comp}, we can obtain the transfer function corresponding to fluctuations in $\delta\phi^{(a)}$,
\begin{align}
    F^{(a)}(\omega)&= 16\beta^2 e^{6\pi}\frac{\omega^2}{(\omega_2^2+\omega^2)^2} \frac{({\omega_2}+{\omega_1\lambda}+2{\omega_1})^2}{(\omega^2+4\omega_2^2)^2\eta_1^4}\nonumber\\
    &\quad\quad\times\bigg(2\omega_2\sin(\frac{3\pi \omega}{2\omega_2}) - \omega\cos\bigg(\frac{3\pi\omega}{2\omega_2}\bigg)\bigg)^2, \label{eq.TF_a}
\end{align}

Similarly, we obtain the transfer function corresponding to $\delta\phi^{(b)}$:
\begin{align}
     F^{(b)}(\omega) &= \frac{32\beta^2\omega_2^4e^{6\pi}(\lambda+1)^2}{(\omega_2^2+\omega^2)^2(\omega^2+4\omega_2^2)^2}\bigg(\frac{1}{\eta_1^2}+\frac{1}{\eta_1\sqrt{2B_{0I}\eta_2}}\bigg)^2\nonumber\\
     &\quad\quad\times\bigg(\omega\sin\bigg(\frac{3\pi\omega}{2\omega_2}\bigg)+2\omega_2\cos\bigg(\frac{3\pi\omega}{2\omega_2}\bigg)\bigg)^2, \label{eq.TF_b}
\end{align}
and the following transfer function corresponding to $\delta\phi^{(c)}$:
\begin{align}
     F^{(c)}(\omega) &= 2\frac{(\beta e^{3\pi})^2}{(\omega^2+4\omega_2^2)^2}\bigg(\frac{\sqrt{2}}{\eta_1\sqrt{B_{0I}\eta_2}}+\frac{1}{\eta_1^2}+\frac{1}{2B_{0I}\eta_2}\bigg)^2\nonumber\\
     &\qquad\quad\times\bigg(\omega\sin\bigg(\frac{3\pi\omega}{2\omega_2}\bigg)+2\omega_2\cos\bigg(\frac{3\pi\omega}{2\omega_2}\bigg)\bigg)^2. \label{eq.TF_c}
\end{align}

\subsection{Computing the Dephasing Rate}\label{sec.4c}
In this subsection, we will compute the dephasing due to white noise using Eq.~\eqref{eq.Gamma_def}. For mathematical simplicity, we shall first compute the contributions of the $F^{(a)}(\omega)$, $F^{(b)}(\omega)$ and $F^{(c)}(\omega)$, separately. To obtain a bound on the total dephasing, we will use an inequality similar to Eq.~\eqref{eq.ineq}. 

To compute a numerical value of the dephasing rate, we shall use the parameter values as mentioned in Table.\ref{tab:stage_params}.
\begin{table}[h!]
\centering
\begin{tabular}{|l|l|}
\hline
\textbf{Parameter} & \textbf{Value} \\ \hline
$\eta_{1} = \eta_{3} = \eta_{5} \equiv \eta_\text{HP}$ & $5 \times 10^{3}~\mathrm{T/m}$ \\
$\eta_{2} = \eta_{5} \equiv \eta_\text{IHP}$ & $1 \times 10^{6}~\mathrm{T/m^2}$ \\
$B_{0H}$ & $1 \times 10^{-3}~\mathrm{T}$ \\
$B_{0I}$ & $0.1~\mathrm{T}$ \\
$T_2 - T_1 = T_4 - T_3$ & $1.5\pi/\omega_2$ \\
\hline
\end{tabular}
\caption{Parameter values used in the five-stage scheme to compute the corresponding dephasing rate and to plot the trajectories in Fig.\ref{fig:B_0-1_trajectory complete approx1micron_latestscheme_nomodellabel}.}\label{tab:stage_params}
\end{table}

For white noise, from Eq.~\eqref{eq.gen_white_noise}, we consider the following pwer spectral density:
\begin{equation}
    S_\text{IHP}(\omega) = A_\text{IHP}^2,
\end{equation} where the subscript denotes the IHP stage. We introduce a dimensionless quantity~\footnote{We see that defining a dimensionless quantity involving the noise parameter will help us obtain an inequality constraining dephasing for the whole scheme, which would involve the noise amplitude in the linear gradient in the harmonic as well as that in the magnetic curvature in the inverted harmonic stages. }
\begin{equation}
    \widetilde{A}_\text{IHP} = \frac{\delta\eta_2}{\eta_2} = A_\text{IHP}\frac{\sqrt{\omega_2}}{\eta_2}.
\end{equation} 
This dimensionless parameter can be interpreted as the ratio between the fluctuation amplitude~\footnote{The square of the fluctuation amplitude can be thought of as the power of the noise, where we integrate the power of the noise at a particular frequency (PSD) over the bandwidth, defined by the transfer function. This was seen to be of the order of $\omega_2$. Hence, $\delta\eta_2 \sim A_\text{IHP}\sqrt{\omega_2}$.} and the signal amplitude of the magnetic field curvature ($\eta_2 = \eta_4$). The dephasing rate accounting only for $F^{(a)}(\omega)$, following Eq.~\eqref{eq.Gamma_def}, is:
\begin{align}
    \Gamma^{(a)}_{2} = 2\int_0^\infty F^{(a)}(\omega) S_\text{IHP}(\omega) \, d\omega.
\end{align}
We don't consider a lower limit on the frequency affecting the system (i.e.,  we take $\omega_{min} = 0$) since the power spectral density is well-behaved throughout the domain and the integral considering the entire spectrum provides an upper bound on the dephasing rate~\footnote{The total experimental time that contributes to the precision of the measurement, and hence the susceptibility to lower noise frequencies, should account for the time duration across all the experimental runs. For instance, say that we need $10^{4}$ experimental runs to obtain the phase information. Then, the lower bound on the frequency in the integral in Eq.~\eqref{eq.Gamma_def}, using the argument in sec.\ref{sec.TFandD} would be $\frac{2\pi}{T_5\times10^{4}}$.}. Now, we compute the dephasing rate explicitly using parameter definitions as Eq.\eqref{eq.def_beta} and Eq.\eqref{eq.def_lambda} for the parameter values provided in Table.\ref{tab:stage_params}:
\begin{align}
    \Gamma^{(a)}_{2} &=4\beta^2 e^{6\pi}\frac{({\omega_2}+{\omega_1\lambda}+2{\omega_1})^2}{\eta_1^4}\frac{ \pi}{9\omega_{2}^{3}}A_\text{IHP}^2\nonumber\\
    &= 2.1\times10^{23}\widetilde{A}^{2}_\text{IHP}\, \text{Hz}
\end{align}

Similarly, the dephasing rate due to only $\delta\phi^{(b)}$ is:
\begin{align}
    \Gamma^{(b)}_{2} &= 16\beta^2e^{6\pi}(\lambda+1)^2\bigg(\frac{1}{\eta_1^2}+\frac{1}{\eta_1\sqrt{2B_{0I}\eta_2}}\bigg)^2\frac{\pi}{9\omega_2}A_\text{IHP}^2\nonumber\\
    &= 5.4\times10^{26}\widetilde{A}^{2}_\text{IHP}\, \text{Hz}
\end{align}

Finally, the dephasing rate accounting only for $\delta\phi^{(c)}$ is:
\begin{align}
    \Gamma^{(c)}_{2} &= \beta^2 e^{6\pi}\bigg(\frac{\sqrt{2}}{\eta_1\sqrt{B_{0I}\eta_2}}+\frac{1}{\eta_1^2}+\frac{1}{2B_{0I}\eta_2}\bigg)^2\frac{\pi}{4\omega_2}A_\text{IHP}^2\nonumber\\
    &= 1.3\times10^{21}\widetilde{A}^{2}_\text{IHP}\, \text{Hz}
\end{align}

The total dephasing for stage 2 of the five-stage process is then bounded by:
\begin{align}
    \Gamma_{2} &\leq \bigg(\sqrt{\Gamma^{(a)}_{2}}+\sqrt{\Gamma^{(b)}_{2}}+\sqrt{\Gamma^{(c)}_{2}}\bigg)^2 \nonumber\\
    &\leq 5.6 \times 10^{26}\widetilde{A}^{2}_\text{IHP} \,\text{Hz}
\end{align}

The dephasing rate in stage 4 of the interferometer is the same. Hence, the total dephasing rate of the two IHP stages (stages 2 and 4) combined is constrained as:
\begin{equation}
    \Gamma^{tot}_\text{IHP} \leq \sqrt{\Gamma_2^2+\Gamma_4^2} = \sqrt{2}\Gamma_2
\end{equation}

Stage 3 lasts for about 0.5 ms, which is negligible compared to other stages. Hence, the dephasing in this stage is neglected. Stages 1 and 5 together form half of a one-loop interferometer under a harmonic potential. Since the noise is modelled using stationary statistics—meaning its statistical properties are invariant in time—the temporal separation between these two stages does not affect the total accumulated dephasing. From ~\cite{HP_Noise}, we know the dephasing rate of a full loop interferometer. As shown in Appendix~\ref{Appendix C}, the total dephasing in the full loop exceeds that of the corresponding half-loop (stages 1 and 5). Therefore, we take the upper bound on the combined dephasing from stages 1 and 5 to be equal to that of the full-loop interferometer, which is given by ~\cite{HP_Noise}:
\begin{align}
    \Gamma_\text{HP}^{tot} \lesssim \frac{8H^2}{\omega_\text{HP}^5} A_\text{HP}^2 \times4.3
\end{align}
where $H = -4 \gamma_e B_{0H} \eta_1 \frac{\chi_\rho}{\mu_0} = 1.8\times10^{10}\,\text{s}^{-1} \text{m}\, \text{kg}^{-1} \text{A}^{-2}$, $\omega_\text{HP} = \omega_{1} = \omega_{5}$ and $A^2_\text{HP}$ is the power spectral density of the white noise considered in the harmonic potential stage. Similar to the definition of the dimensionless quantity related to the noise amplitude in the case of the IHP, we define:
\begin{equation}
    \widetilde{A}_\text{HP} = \frac{\delta\eta_1}{\eta_1} = A_\text{HP}\frac{\sqrt{\omega_{1}}}{\eta_{1}}.
\end{equation}
This dimensionless parameter can be interpreted as the ratio between the fluctuation amplitude and the signal amplitude of the magnetic field gradient in the harmonic potential stage.
From this, we get 
\begin{align}
    \Gamma_\text{HP}^{tot} \lesssim 1.2\times 10^{7} \widetilde{A}_\text{HP}^2\,\, \text{Hz}
\end{align}
Hence, the total dephasing rate is:
\begin{align}
    \Gamma^{tot} &\leq \bigg(\sqrt{\Gamma^{tot}_\text{IHP}}+\sqrt{\Gamma_\text{HP}^{tot}}\bigg)^2 \nonumber\\
    &\sim \bigg(2.8\times10^{13}\widetilde{A}_\text{IHP}+3.5\times10^{3}\widetilde{A}_\text{HP}\bigg)^2 \, \text{Hz}\label{eq.Gamma_tot_final}
\end{align}

\subsection{Obtaining bounds on the parameter}
One of the measures of coherence \cite{PhysRevLett.113.140401} is given by:
\begin{equation}
    \text{Coherence} = e^{-\Gamma\tau} \in [0,1]\label{eq.cohdef}
\end{equation}
where $\Gamma$ is the dephasing rate and $\tau$ is the experimental time of this stage for one run of the total experiment~\footnote{We can intuitively understand this quantification of coherence by looking at its behaviour for asymptotic values of the dephasing rate $\Gamma$. As $\Gamma \to 0$, $e^{-\Gamma\tau} \to 1$ and as $\Gamma \to \infty$, $e^{-\Gamma\tau} \to 0$. Thus we see that $e^{-\Gamma\tau}$ decreases with increasing dephasing rate, being a good measure of coherence.}. We shall constrain the decoherence to at most 90\%, i.e., we will now find the bound on the parameters $A_\text{IHP}$ and $A_\text{HP}$, such that the noise utmost decoheres the system to 10\% coherence, where coherence is defined by Eq.~\eqref{eq.cohdef}. We assume the parameter values as considered in sec.\ref{sec.4c}, then $\tau = T_5 = 0.31$s. 
\begin{align}
    \text{Coherence} &= e^{-\Gamma\tau} \geq 0.1 \nonumber\\
    \implies & \Gamma \leq 7.4 \,\text{Hz} \label{eq.bound_gamma}
\end{align}


From Eq.~\eqref{eq.Gamma_tot_final} and Eq.~\eqref{eq.bound_gamma}, we can obtain the bound on the parameter, $A_\text{IHP}$ and $A_\text{HP}$, 
\begin{align}
   &\bigg(2.8\times10^{13}\widetilde{A}_\text{IHP}+3.5\times10^{3}\widetilde{A}_\text{HP}\bigg)^2 \, \text{Hz} \leq 7.4 \, \text{Hz} \nonumber\\
    &\implies 0.8\times10^{10}\widetilde{A}_\text{IHP}+\widetilde{A}_\text{HP} \lesssim 0.78\times10^{-3} \label{eq.ineqA}
\end{align}
Consider the above equation. The RHS defines the order of magnitude constraining the upper bound on $\widetilde{A}_\text{HP}$. Hence, if we could constrain to $\order{10^{-6}}$, then it can be neglected in comparison to the RHS. This would correspond to a constraint on the physical value of the noise parameter, $A_\text{HP}$, as:
\begin{align}
    \widetilde{A}_\text{HP}&\lesssim\order{10^{-6}} \label{eq.tilde_AHP_bound}\\
    \implies A_\text{HP}&\lesssim\order{10^{-4}} \, \text{T}\,\text{m}^{-1}\,\text{Hz}^{-1/2}.
\end{align}
Compared to the bound laid on the white noise parameter in the harmonic potential stage in the Ref.\cite{HP_Noise}, the above bound is looser, and the constraint on the harmonic potential stage is achievable. Now we can obtain a bound on $\widetilde{A}_\text{IHP}$ after neglecting $\widetilde{A}_\text{HP}$ in Eq.\eqref{eq.ineqA}: 
\begin{align}
    \widetilde{A}_\text{IHP} &\lesssim 0.98\times10^{-13}\label{eq.tilde_AIHP_bound}\\
    \implies A_\text{IHP} &\lesssim 1.7\times10^{-8} \, \text{T}\,\text{m}^{-2}\,\text{Hz}^{-1/2}
\end{align}
In Table.\ref{tab:stage_params}, since we consider the same value for the gradients in each of the harmonic potential stages, we shall denote it as $\eta_\text{HP}$. Similarly, since the values of the magnetic field curvature considered in stages 2 and 4 are the same, we denote it as $\eta_\text{IHP}$. Recall that the dimensionless quantities $\widetilde{A}_\text{HP}$ and $\widetilde{A}_\text{IHP}$ can be interpreted as the ratio of the fluctuations in the gradient in the harmonic and curvature to their signal strengths, respectively. Hence, we also obtain a bound on the noise fluctuations from Eq.\eqref{eq.tilde_AHP_bound} and Eq.\eqref{eq.tilde_AIHP_bound} as:
\begin{align}
    \frac{\delta\eta_\text{HP}}{\eta_\text{HP}} &\lesssim 10^{-6} \\
    \frac{\delta\eta_\text{IHP}}{\eta_\text{IHP}} &\lesssim 10^{-13}
\end{align}

We now compare the bounds obtained above on the magnetic field gradient and the curvature by constraining the dephasing rate to those obtained by constraining the contrast of the interferometer as analysed in Ref.\cite{PhysRevA.111.052207}. In Ref.\cite{PhysRevA.111.052207}, the contrast was computed for fluctuations in the magnetic field gradient in the harmonic potential stages and fluctuations in the magnetic field curvature in the inverted harmonic potential stages in the five-stage process, where the first harmonic potential stage ends with zero velocity and maximum spatial separation of the two arms of the interferometer possible in this stage. Constraining the contrast as $\mathcal{C}(t) \geq 0.9$, the bound obtained on $\delta\eta_{IHP}/\eta_{IHP}$, where $\eta_{IHP}$ is the magnetic field curvature in the inverted harmonic potential stage, is: $\delta\eta_{IHP}/\eta_{IHP}\leq \order{10^{-6}}$, and the bound on the magnetic gradient fluctuations is slightly lower than that of $\delta\eta_{HP}/\eta_{HP}\leq \order{10^{-6}}$. Comparing this with the bounds obtained above, by constraining the dephasing rate, we find that the dephasing rate-based criterion imposes a notably tighter limit on fluctuations of the magnetic field curvature, $\eta_{IHP}$, whereas the bound on the magnetic field gradient is of a similar order of magnitude. This shows that in our scheme, constraining the dephasing rate is sufficient to ensure a good contrast. This is consistent with Ref.~\cite{HP_Noise}, where applying a dephasing constraint in a purely harmonic-potential interferometer leads to negligible contrast loss. The intuition behind the observed (seven-order-of-magnitude) disparity for the curvature bounds is that contrast depends only on the net end-point displacement of the two trajectories, whereas the dephasing rate is sensitive to the entire path history. Consequently, even when the overall displacement is small—implying little contrast degradation—the accumulated phase variance from trajectory deviations along the full path can still be large. Now we will explicitly analysis the Humpty-Dumpty problem in inverted harmonic oscillators.

\section{Humpty-Dumpty Analysis}

We find that it is informative to look at superposition schemes with shorter experimental durations, but where the interferometer may not close completely, or the relative velocity of the two arms at the closure of the interferometer is nonzero, as motivated by the remarks at the end of Sec.\ref{sec.IHP-5stage}. Hence, we would like to probe how far these conditions can be relaxed so that we can consider these schemes as well. To determine the constraints on the allowable classical differences in position (\( \Delta x \)) and momentum (\( \Delta p \)) between the two arms, the quantum uncertainties in these observables (\( \delta x \)) and (\( \delta p \)) must be larger than the corresponding classical values\cite{Schwinger1988}.

We consider the initial state to be described by a Gaussian-shaped wave packet (GSWP). The general form of a Gaussian-shaped wavepacket (GSWP) can be written as:
\begin{equation}
\psi(x, t=0) = N_0 \exp\left[ -\frac{(x - x_0)^2}{2\sigma_0^2} + i \bigg(\frac{a_0}{2} x^2 + b_0 x + c_0\bigg) \right]
\end{equation}
where \( N_0 \) is the normalization factor, \( \sigma_0 \) is the initial wavepacket width, \( x_0 \) is the center position of the wavepacket, and \( a_0 \), \( b_0 \), and \( c_0 \) are phase-related parameters. A GSWP remains a GSWP under time evolution in both a harmonic potential (HP) and an inverted harmonic potential (IHP), thus:
\begin{equation}
\psi(x, t) = N_t \exp\left[ -\frac{(x - x_c(t))^2}{2\sigma_x^2(t)} + i\bigg( \frac{a_t}{2} x^2 + b_t x + c_t\bigg) \right]
\end{equation}
\( x_c(t) \) is the classical equation of motion of the wave packet, and all parameters with a subscript $t$ is time-dependent.


In general, the initial state of the particle in an interferometer is likely to be taken to be a minimum uncertainty Gaussian wavepacket (MUWP), since we want the least uncertainty in both quantities and not compromise one for the other in the case of an interferometer. Thus, by understanding the spread of the MUWP state, we can obtain a lower bound on the position uncertainty. For an MUWP state, the initial parameters are:
\begin{equation}
N_0 = \frac{1}{\sqrt{2\pi} \,\sigma_0}, \quad
a_0 = 0, \quad
b_0 = \frac{p_0}{\hbar}, \quad
c_0 = -\frac{p_0 x_0}{\hbar} \nonumber
\end{equation}
where $p_0$ is the initial momentum. The width as a function of time for the harmonic is~\cite{PhysRevA.111.052207}: 
\begin{equation}
\sigma_x^{\text{H}}(t) = \sigma_0 \left( \frac{\hbar^2}{4 m^2 \omega^2 \sigma_0^4} \sin^2(\omega t) + \cos^2(\omega t) \right)^{1/2} \label{eq.sigma_H}
\end{equation}
And that for the inverted harmonic potential is~\cite{PhysRevA.111.052207}:
\begin{equation}
\sigma_x^{\text{I}}(t) = \sigma_0 \left( \frac{\hbar^2}{4 m^2 \omega^2 \sigma_0^4} \sinh^2(\omega t) + \cosh^2(\omega t) \right)^{1/2} \label{eq.sigma_I}
\end{equation}
For reference, the wavepacket evolution of a free particle is given by: \begin{equation}
\sigma_x^{\text{free}}(t) = \sigma_0 \left( 1 + \frac{\hbar^2t^2}{4 m^2\sigma_0^4} \right)^{1/2} \label{eq.sigma_free}
\end{equation}

\begin{figure}
    \centering
    \includegraphics[width=\linewidth]{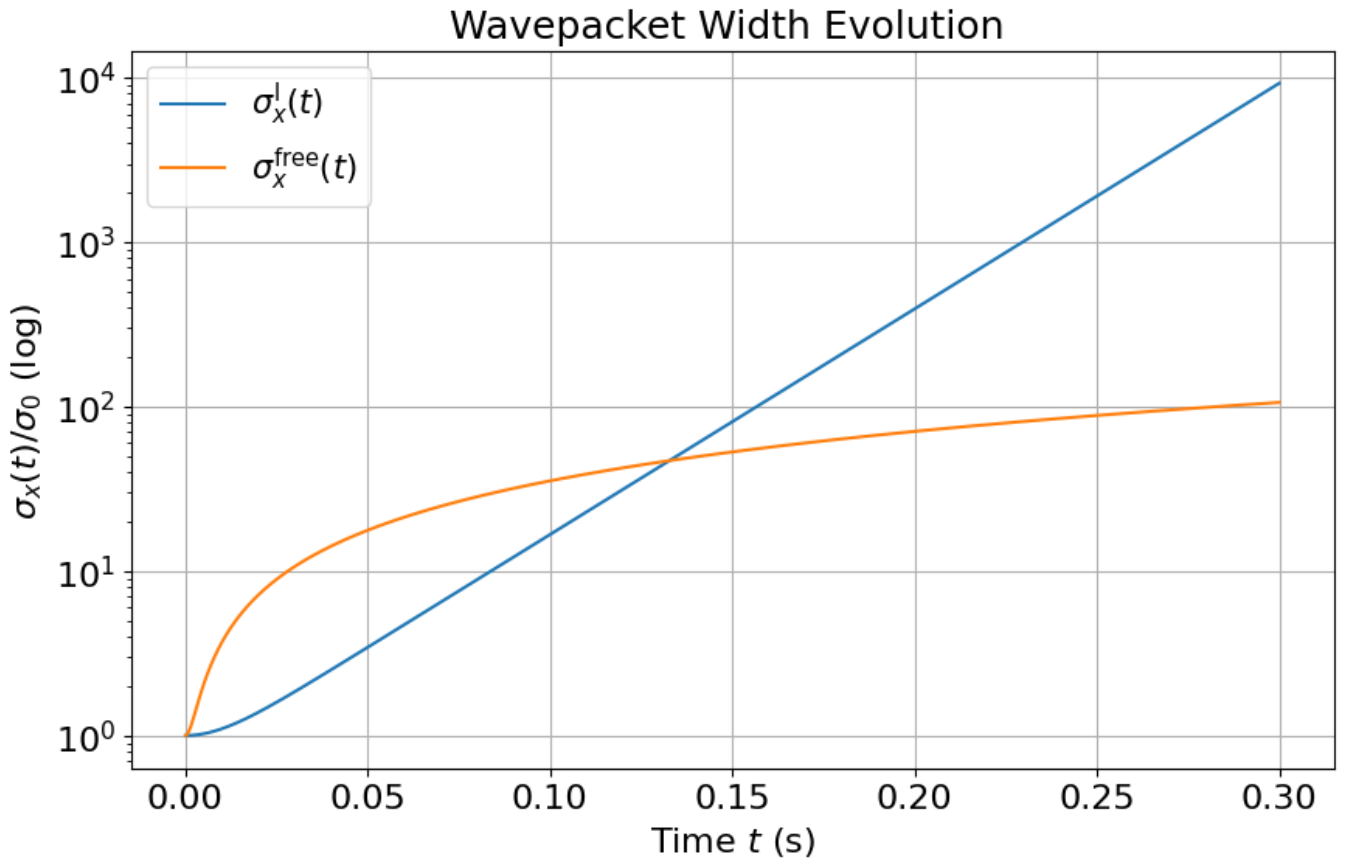}
    \captionsetup{justification=RaggedRight}
    \caption{Evolution of the width of a minimum uncertainty Gaussian wavepacket: The blue graph depicts the evolution under an IHP potential: $\eta_I = 1\times 10^6$\,Tm$^{-2}$ and $B_{0I}=0.1$\,T, and the orange graph depicts evolution under free evolution.}
    \label{fig:wavepacket_evolution}
\end{figure}

Now we shall apply eq.\ref{eq.sigma_H} and eq.\ref{eq.sigma_I} to each of the stages, with the ultimate goal of estimating the order of magnitude of the width of the wavepackets at the end of the interferometer. In our experiment, we shall assume that the particle is in a motionally cooled state, i.e., it is in the ground state of the harmonic potential in the initial stage. Then the initial width of the wavepacket is given by:
\begin{equation}
    \sigma_0 = \sqrt{\frac{\hbar}{2m\omega_1}}
\end{equation}
where $\omega_1$ is the natural frequency of the harmonic potential in the first stage. For $\eta_1 = 5\times 10^3$ T m$^{-1}$, the corresponding natural frequency for a particle of mass $10^{-15}$ kg is, $\omega_1 = 354$Hz. Hence, the initial width at the beginning of stage 1 is
\begin{equation}
    \sigma_{01} = \sqrt{\frac{\hbar}{2m\omega_1}} \sim 1.2\times10^{-11} \, \text{m} \label{eq.width_s1}
\end{equation}
An MUWP width doesn't change under a harmonic potential, hence, the initial width at the beginning of stage 2 is the same as given by eq.\ref{eq.width_s1}. Stage 3 doesn't affect the order of magnitude of the width, as can be seen from the trigonometric functions involved in eq.\ref{eq.sigma_H}, and the short duration of the HP stage in comparison to the IHP stage. Hence, the width can be assumed to be a function of time evolving according to eq.\ref{eq.sigma_I} with initial width given by eq.\ref{eq.width_s1}. The evolution of the wavepacket is shown in Fig.\ref{fig:wavepacket_evolution}.

We observe from the figure that though the free evolution broadens the wavepacket fast initially, in the timescale considered, the wavepacket broadens much more under the inverted harmonic potential. In the five-stage process, at the end of the interferometer, the size of the wavepacket becomes about $10^4$ times the initial width. Hence, at the end of the interferometer, the width of the GSWP is given by
\begin{equation}
    \sigma_{05} \sim \order{10^{-7}} \, \text{m} \label{eq.widths5}
\end{equation}

To obtain the tolerance in deviations in momentum at the end of the interferometer, we need a lower bound on the quantum uncertainty of the momentum wavefunction at the end of the interferometer. Though an MUWP need not develop into an MUWP under an inverted harmonic potential, the lower bound on the momentum uncertainty can be obtained by imposing the minimum uncertainty at the end of the interferometer:
\begin{equation}
    \sigma_{05}^p\geq\frac{\hbar}{2\sigma_{05}} \sim 0.5\times10^{-27} \text{kg m s}^{-1}
\end{equation}

Intuitively, we can see that as long as at the end of the interferometer, the difference in the most probable positions of the wavefunctions is less than $\order{10^{-7}}$m and if the classical deviation in momentum is less than the quantum uncertainty, we can conduct an interference experiment, as there would be sufficient overlap of the wavefunctions. We also note that the tolerance to momentum deviations at the end of the interferometer may be a strong constraint, and hence we should focus on closing the interferometer with each state near zero velocity and can compromise on the precision of spatial overlap. A more accurate bound can be obtained from the uncertainty relation mentioned in ~\cite{cjp-2014-0553}. 

To verify in a mathematical manner, one could compute the contrast at the end of the interferometer. Contrast is a measure of the visibility of interference fringes, accounting for the loss in visibility due to the lesser overlap of the interfering wavefunctions. It lies in the range [0,1], where $\mathcal{C} = 1$ means that the interferometer closes and $\mathcal{C}=0$ means that the interferometer doesn't close and hence spin-readout is not possible. The contrast is given by~\cite{Schwinger1988}:
\begin{equation}
    \mathcal{C}(t) = \exp\bigg(-\frac{1}{2}\bigg[\bigg(\frac{\Delta x(t)}{\sigma_x}\bigg)^2+\bigg(\frac{\Delta p(t)}{\sigma_p}\bigg)^2\bigg]\bigg) \label{eq.Contrast}
\end{equation}
where $\Delta x(t)$ and $\Delta p(t)$ are the differences in position and momentum between the two arms of the interferometer, at time $t$. This has to be computed at the end of the interferometer to understand if the interference can be carried out.

\section{Conclusion}

In this work, we proposed a modification to the five-stage protocol for generating macroscopic spatial superpositions in a spin-1 quantum system, such as a nitrogen vacancy (NV) centre in a nanodiamond. Our scheme builds upon the proposal in Ref.~\cite{PhysRevA.111.052207}, which employs inverted harmonic potentials (IHPs) to accelerate wave-packet separation and achieve a target superposition size of approximately \(1~\mu\mathrm{m}\). By terminating the initial separation stage at the point of maximum velocity, rather than maximum spatial separation, the subsequent IHP stage drives a faster expansion of the superposition, which differs markedly from Ref.~\cite{PhysRevA.111.052207}. For fixed experimental parameters, our approach increases the achievable separation by nearly an order of magnitude within the same time duration (0.3 s; see Appendix~\ref{Appendix B}). Thus, the present scheme achieves the same superposition size (\(1~\mu\mathrm{m}\)) in a shorter time, providing a significant advantage in mitigating decoherence.

We derived the transfer functions for the IHP stage, which allow quantitative evaluation of dephasing due to white noise in the magnetic-field gradients ($\partial B/\partial x$) and curvatures ($\partial^2 B/\partial x^2$). Starting from the system’s Lagrangian with fluctuating magnetic fields, we computed the induced phase fluctuations and, using the Wiener–Khinchin theorem~\cite{Wiener:1930,Khintchine1934}, expressed the resulting dephasing rate in terms of transfer functions and the noise power spectral density.

Our analysis shows that dephasing arises both directly from fluctuations in the magnetic field (gradient and curvature) and indirectly through noise-induced variations in the velocity and trajectories of the interferometer arms. We find that the dominant contribution comes from the IHP stages. We consider white noise statistics, i.e. noise with flat power spectral density (PSD), for both the HP and IHP stages. For the dephasing rate in the harmonic potential stage, we use the results obtained in Ref.\cite{HP_Noise}. We find that to preserve at least \(10\%\) coherence over the full protocol (0.3 s), the magnetic-field curvature noise amplitude (defined as $\sqrt{\text{PSD}}$) must satisfy \(A_\text{IHP} \lesssim 1.7\times 10^{-8}~\mathrm{T\,m^{-2}\,Hz^{-1/2}}\), and the gradient noise amplitude must satisfy \(A_\text{HP} \lesssim 10^{-3}~\mathrm{T\,m^{-1}\,Hz^{-1/2}}\). These bounds correspond to an upper limit on the dephasing rate of \(\Gamma \leq 7.4~\mathrm{Hz}\). The noise-to-signal ratio permitted in the magnetic field gradient and curvature are $10^{-6}$ and $10^{-13}$, respectively. This highlights the stringent magnetic field stability demanded during the IHP stage. Further, we note that within the IHP stage, the trajectory fluctuations contribute the most to the dephasing rate, followed by the velocity fluctuations, and the direct effect of the magnetic field curvature fluctuations is relatively low.

We also discussed the Humpty–Dumpty problem, where fluctuations in the magnetic field can prevent perfect closure of the interferometer trajectories. A key advantage of the IHP stage is that the wavefunction broadens exponentially with time, unlike free evolution, where broadening is linear. Over the 0.3 s protocol, the IHP broadens the wavepacket by nearly two orders of magnitude more than free evolution, yielding a final wavepacket width of order \(10^{-7}\,\mathrm{m}\). Provided trajectory fluctuations remain below this scale and the interferometer closes at near-zero relative velocity, the resulting contrast remains close to unity. Thus, position fluctuation-induced loss of contrast is negligible under these conditions.

Additionally, we find that the bounds obtained on the noise parameters are stricter if we constrain the dephasing rate, as we have done in this paper, as compared to the bounds obtained by constraining the contrast, as was done in Ref.\cite{PhysRevA.111.052207}. A similar trend was found in Ref.\cite {HP_Noise}: trajectory fluctuations can contribute non-negligibly to the dephasing rate but remain low enough for the contrast to remain high.  In future, we will also do the noise analysis for the two-dimensional interferometer, and include the switching profile from going from one magnetic profile to another.

Overall, these findings provide quantitative constraints to guide experimental design for quantum-gravity–motivated entanglement tests and related macroscopic quantum experiments. Future work could explore protocols where interferometer closure is deliberately imperfect, with only partial wavefunction overlap at recombination, as well as tolerance to finite transition times between stages and their impact on interferometric contrast.

\section*{Acknowledgements}
AM’s research is funded in part by the Gordon and Betty Moore Foundation through Grant GBMF12328, DOI10.37807/GBMF12328. AM’s work is also supported by the Alfred P. Sloan Foundation under Grant No. G-2023-21130. SNM would like to thank the organisers of the workshop, Schr\"odinger Cats: The Quest to find the edge of the quantum world, held at OIST, Japan, for facilitating academic interactions that led to this collaboration. SNM acknowledges support from the Kishore Vaigyanik Protsahan Yojana (KVPY) fellowship, SX-2011055, awarded by the Department of Science and Technology, Government of India.

\bibliography{References}
\appendix
\section{Quantization axis and effective 1D potential}\label{Appendix A}
The spin dynamics in one axis are inherently coupled to the other axes because the spin operators for different directions do not commute. To decouple the spin dynamics along a chosen axis, a strong, static magnetic field is applied along that direction. When the Hamiltonian \( H \) commutes with the spin operator along the chosen axis, say \( S_x \), i.e., \([H, S_x] = 0\), the expectation value \( \langle S_x \rangle \) becomes a conserved quantity, as dictated by the Heisenberg equation: 
\begin{equation}
\frac{d\langle S_x \rangle}{dt} = \frac{i}{\hbar} \langle [H, S_x] \rangle = 0.
\end{equation}
This effectively decouples the x-axis dynamics from the other two components. The transverse components \( \langle S_y \rangle \) and \( \langle S_z \rangle \) precess about the x-axis at the Larmor frequency, \(\omega_L = \gamma_e B_0\), where \(\gamma_e / 2\pi \approx 28 \, \text{GHz/T}\) is the electron gyromagnetic ratio and \( B_0 \) is the bias field strength. This defines the x-axis as the spin quantisation axis, ensuring that the spin states \( |m_x = \pm 1, 0\rangle \) are eigenstates of the dominant Hamiltonian term, and hence the spatial evolution is dominated by dynamics in this direction \cite{fonsecaromero2024}.

In the QGEM protocol, we work with a nitrogen-vacancy (NV) centre in a diamond nanoparticle. Without a defined quantisation axis along the x-axis, the NV centre’s intrinsic zero-field splitting (\( D S_{NV}^2 \), with \( D \approx 2.87 \, \text{GHz} \)) could cause unwanted precession, disrupting the superposition. A bias magnetic field \( B_{0} = 10^{-3} \, \text{T} \) gives $\frac{\gamma_e}{2\pi}B_{0} = 2.8\times10^{-2}$GHz. This value is smaller than $D$. Hence, for precession along the x-axis, we should align the NV axis along the direction of the bias magnetic field as much as possible.

Additionally, we do not want free evolution along the y and z axes; we want control over the y and z axis positions of the particle. Hence, we apply a strong trap along these directions. For the x-direction to still define the quantisation axis, we have to estimate the lower bound on the magnitude of the bias field required along the x-axis such that the gradients forming the trap along the other directions give rise to magnetic fields sufficiently smaller than the bias field along the x-axis. Since we are aiming for a steeper trap along the y and z axes, let the gradient along these directions be of the order of $3\times10^4$T m$^{-1}$, which corresponds to a natural frequency of about 2100Hz. We want the position states along these axes to be in the motional cooled state - the ground state of the corresponding harmonic potential. Hence, the spread of the ground state wavefunction along these axes would be:
\begin{equation}
    \Delta y = \Delta z = \sqrt{\frac{\hbar}{2m\omega_y}} = 5\times 10^{-12}\,\text{m}
\end{equation}
The maximum magnetic field achieved in this spatial interval is $\eta_y\times\Delta y = 1.5\times10^{-7}$ T. Hence, the assumed bias field of $10^{-3}$T, is very much sufficient to define the spin quantisation axis.

\section{Comparison} \label{Appendix B}
To compare the influence of the final velocity at the end of the initial stage, we consider two models:
\begin{itemize}
    \item \textbf{Model I}: The first stage ends with zero velocity.
    \item \textbf{Model II}: The first stage ends with maximum velocity.
\end{itemize}

\begin{figure}[h!]
    \centering
    \begin{subcaptionbox}{Model I\label{fig:model_1}}[0.5\textwidth]
        {\includegraphics[width=\linewidth]{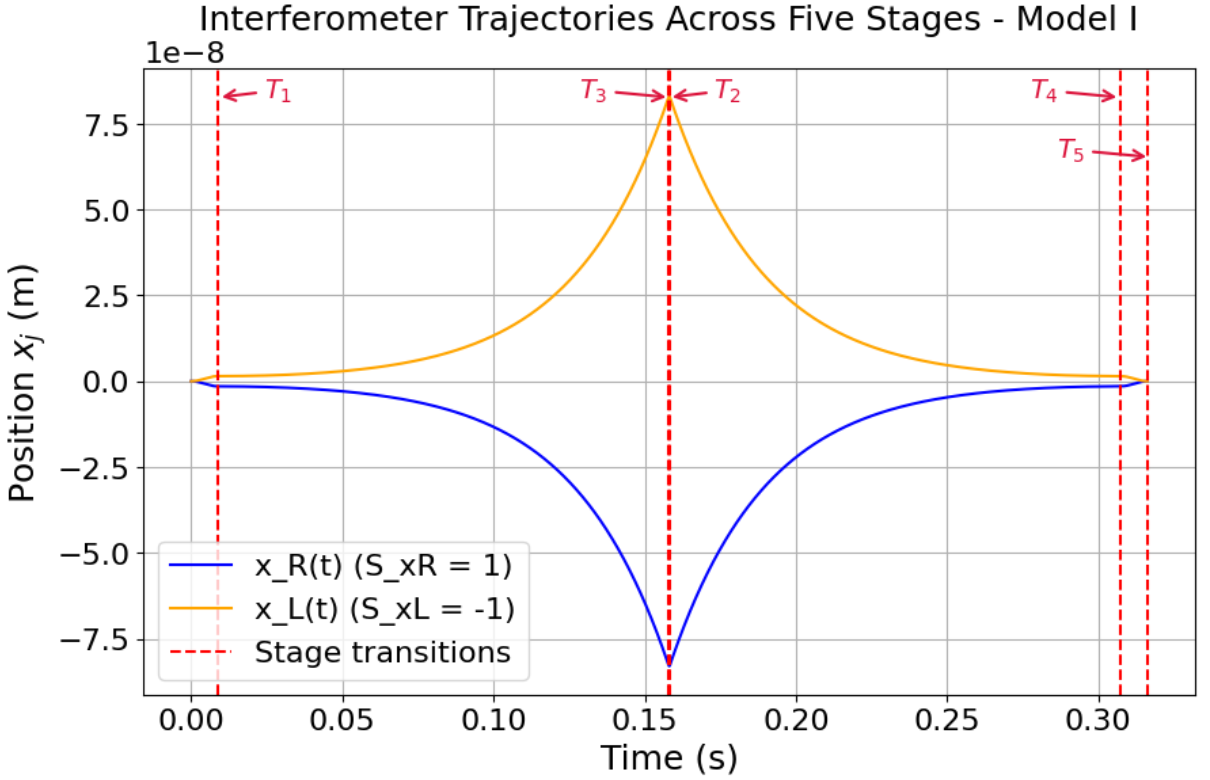}}
    \end{subcaptionbox}
    \hspace{0.05\textwidth}
    \begin{subcaptionbox}{Model II\label{fig:model_2}}[0.5\textwidth]
        {\includegraphics[width=\linewidth]{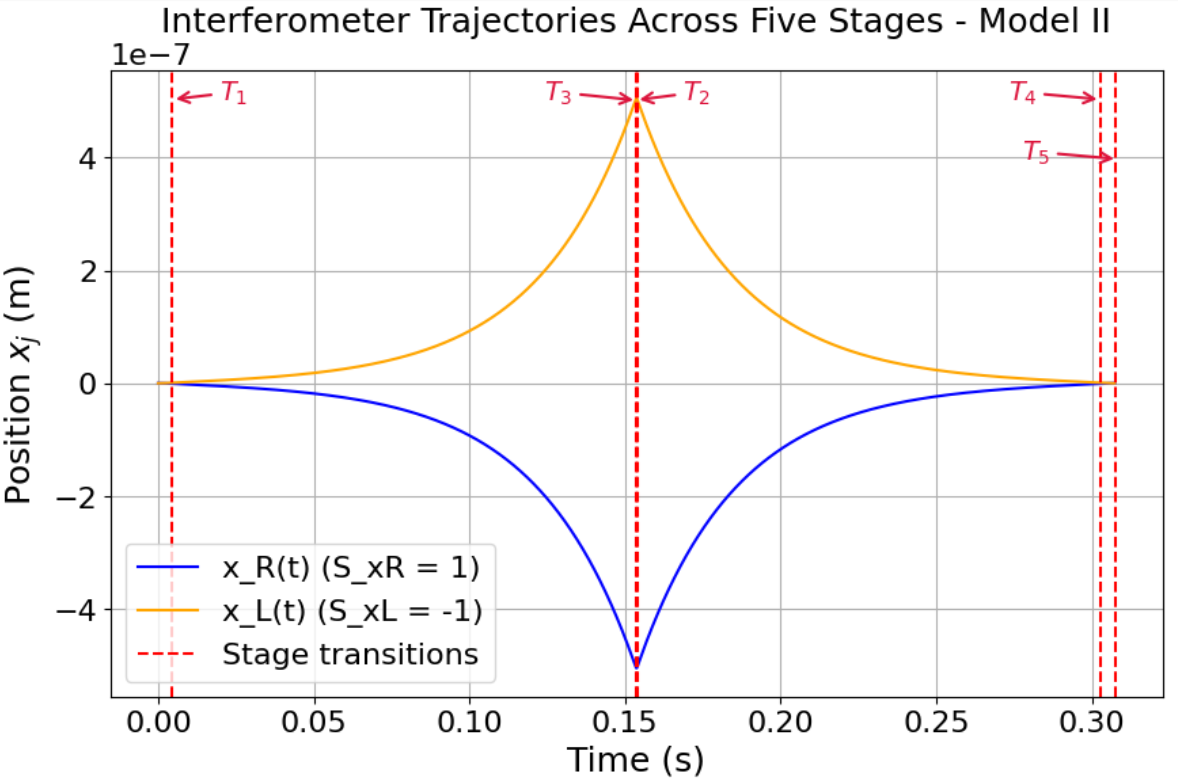}}
    \end{subcaptionbox}
    \captionsetup{justification=RaggedRight}
    \caption{Comparing Model I and Model II: In Model I, the end of the harmonic stage is when the two spin states have maximum spatial superposition achievable in the harmonic potential. In Model II, the end of the harmonic stage is when the two spin states have maximum velocity, achievable in the harmonic potential. The latter helps give an initial velocity at the beginning of the inverted harmonic potential stage. We see that the second model gives us a superposition size higher than the first model by one order of magnitude, within a similar time duration.}\label{fig:combined}
\end{figure}

Now we simulate the two models with the system parameters for both taken as the same according to Table.\ref{tab:stage_params}. $B_{0H}$ can be taken to be zero since it doesn't affect the superposition size. The results of this analysis (with the interferometer trajectories compared in Fig.\ref{fig:combined}) are presented in Table~\ref{tab:comparison}. It shows that Model II completes the protocol in a slightly shorter duration and achieves a significantly larger superposition size, making it the more favourable configuration under the specified parameters.

\begin{table}[ht!]
\caption{\label{tab:comparison}Comparison of the two models under identical parameter settings.}
\begin{ruledtabular}
\begin{tabular}{lcc}
Quantity & Model I & Model II \\
\hline
Total duration (s) & 0.316 & 0.307 \\
Superposition size (m) & $1.7 \times 10^{-7}$ & $1 \times 10^{-6}$ \\
\end{tabular}
\end{ruledtabular}
\end{table}

Thus, the time duration to get a one micron superposition employing model I scheme, using the same system parameters as in Table.\ref{tab:stage_params} would be higher. We observe that the required increase is about 0.13s. Thus, the exponential that appears in the transfer function (Eqs.\eqref{eq.TF_a}-\eqref{eq.TF_c}), which shall appear in the transfer function of the model I as well, would increase the susceptibility of the system to noise. Also, since the duration of the experiment is higher in model I, the bound on the dephasing rate would be stronger. Hence, we expect the order of magnitude of bounds on the system parameters in model I not to be better than model II.

\section{Noise - Harmonic one loop vs two quarter loops}\label{Appendix C}
The fluctuations in the phase in case of a harmonic potential landscape with a spin superposition can be mathematically depicted as:
\begin{align}
    \delta \phi = \int_0^T \Bigg[&-\frac{1}{2}m\bigg(-\frac{\chi_\rho}{\mu_0}\bigg)2\eta_0\delta\eta(t)(x_R^2-x_L^2) \nonumber \\
    &- \frac{1}{2}m\bigg(-\frac{\chi_\rho}{\mu_0}\bigg)\eta_0^2(2x_R\delta x_R-2x_L\delta x_L) \nonumber\\
    &- \hbar\gamma_e\delta\eta(t) (x_R+x_L) - \hbar\gamma_e\eta_0 (\delta x_R+\delta x_L) \nonumber \\
    &+ \frac{ \chi_\rho m}{\mu_0}B_0\delta\eta(t)(x_R-x_L)
    \nonumber \\
    &+ \frac{ \chi_\rho m}{\mu_0}B_0\eta_0(\delta x_R-\delta x_L)\Bigg]\,dt \label{eq.Effmag_pert_action}
\end{align}
where the notation is the same as in \ref{sec.3A}.

The only difference between a complete loop and a half loop (stage 1 + stage 5 of the five-stage process) is accounted for by the position and velocity quantities and the time of integration. In comparison, the difference and the sum of positions in the complete loop are higher as compared to those in the half loop. Hence, the noise analysis conducted for the full loop will provide an upper bound for the noise in the half loop.

\end{document}